\documentclass[a4paper,12pt]{article}
\usepackage{a4,amsmath,amssymb,axodraw,slashed,epsfig}
\usepackage[square,comma,numbers,sort&compress]{natbib}

\begin{document}


\newcommand{\bed}{\begin{displaymath}}
\newcommand{\eed}{\end{displaymath}}
\newcommand{\beq}{\begin{equation}}
\newcommand{\eeq}{\end{equation}}
\newcommand{\bea}{\begin{eqnarray}}
\newcommand{\eea}{\end{eqnarray}}
\newcommand{\tgb}{{\rm tg}\beta}
\newcommand{\tga}{{\rm tg}\alpha}
\newcommand{\stgb}{{\rm tg}^2\beta}
\newcommand{\sia}{\sin\alpha}
\newcommand{\coa}{\cos\alpha}
\newcommand{\sib}{\sin\beta}
\newcommand{\cob}{\cos\beta}
\newcommand{\MS}{\overline{\rm MS}}
\newcommand{\st}{\tilde{t}}
\newcommand{\sgl}{\tilde{g}}
\newcommand{\Dmb}{\ensuremath{\Delta_b}}

\newcommand{\tb}{{\rm tg}\beta}
\newcommand{\gluino}{\widetilde{g}}
\newcommand{\squark}{\tilde{q}}
\newcommand{\Mq}[1]{m_{\tilde{q}_{#1}}}
\newcommand{\mq}{m_{q}}
\newcommand{\Aq}{A_q}
\newcommand{\mix}{\widetilde{\theta}}
\newcommand{\Lag}{\mathcal{L}}
\newcommand{\ls}{\lambda_{s}}
\newcommand{\lb}{\lambda_{b}}
\newcommand{\lt}{\lambda_t}
\newcommand{\Deltamb}{\Delta_b}
\newcommand{\DeltambQCD}{\Delta_b^{QCD}}
\newcommand{\DeltambELW}{\Delta_b^{elw}}
\newcommand{\mb}{m_{b}}
\newcommand{\mt}{m_{t}}
\newcommand{\bR}{b_R}
\newcommand{\bL}{b_L}
\newcommand{\sR}{s_R}
\newcommand{\sL}{s_L}
\newcommand{\At}{A_t}
\newcommand{\Sb}{\Sigma_b}
\newcommand{\Ao}[1]{A_0(#1)}
\newcommand{\Co}[5]{C_0(#1,#2;#3,#4,#5)}
\newcommand{\as}{\alpha_s}
\newcommand{\Mg}{m_{\tilde{g}}}
\newcommand{\Mb}[1]{m_{\tilde{b}_{#1}}}
\newcommand{\Mt}[1]{m_{\tilde{t}_{#1}}}
\newcommand{\NL}{\nonumber\\}
\newcommand{\bs}{\widetilde{b}}
\newcommand{\str}{\widetilde{s}}
\newcommand{\bsL}{\bs_L}
\newcommand{\bsR}{\bs_R}
\newcommand{\strL}{\str_L}
\newcommand{\strR}{\str_R}
\newcommand{\ts}{\widetilde{t}}
\newcommand{\tsR}{\widetilde{t}_R}
\newcommand{\tsL}{\widetilde{t}_L}
\newcommand{\higgsino}{\widetilde{h}}
\newcommand{\stopx}{\tilde{t}}
\newcommand{\sbottom}{\tilde{b}}
\newcommand{\strange}{\tilde{s}}
\newcommand{\gh}{g_b^h}
\newcommand{\gH}{g_b^H}
\newcommand{\gA}{g_b^A}
\newcommand{\ght}{\tilde{g}_b^h}
\newcommand{\gHt}{\tilde{g}_b^H}
\newcommand{\gAt}{\tilde{g}_b^A}
\newcommand{\stb}{{\rm tg}^2\beta}
\newcommand{\Order}[1]{{\cal{O}}\left(#1\right)}
\newcommand{\Msusy}{M_{SUSY}}
\newcommand{\gluon}{g}
\newcommand{\sttop}{\widetilde{t}}
\newcommand{\CF}{C_F}
\newcommand{\CA}{C_A}
\newcommand{\TR}{T_R}
\newcommand{\Tr}[1]{{\rm Tr}\left[#1\right]}
\newcommand{\dk}{\frac{d^nk}{(2\pi)^n}}
\newcommand{\dqq}{\frac{d^nq}{(2\pi)^n}}
\newcommand{\partderiv}[1]{\frac{\partial}{\partial#1}}
\newcommand{\kmu}{k_{\mu}}
\newcommand{\qmu}{q_{\mu}}
\newcommand{\eUV}{\epsilon}
\newcommand{\dMq}[1]{\delta\Mq{#1}}
\newcommand{\gs}{g_s}
\newcommand{\MSbar}{\overline{\rm MS}}
\newcommand{\gPhit}{\tilde{g}_b^{\Phi}}
\newcommand{\muR}{\mu_{R}}
\newcommand{\bbbar}{b\bar{b}}
\newcommand{\MPhi}{M_{\Phi}}
\newcommand{\dlt}{\delta\lambda_t}
\newcommand{\GF}{{\rm G_F}}
\newcommand{\gPhi}{g_b^{\Phi}}
\newcommand{\mbMS}{\overline{m}_{b}}
\newcommand{\asrun}[1]{\as(#1)}
\newcommand{\NF}{N_F}
\newcommand{\gtPhi}{g_t^{\Phi}}
\newcommand{\Log}[1]{\log\left(#1\right)}
\newcommand{\smallaeff}{small $\alpha_{eff}$}
\newcommand{\ra}{\rightarrow}
\newcommand{\tautau}{\tau^+\tau^-}

\newcommand{\Quark}[4]{\ArrowLine(#1,#2)(#3,#4)}
\newcommand{\Gluino}[5]{\Gluon(#1,#2)(#3,#4){3}{#5}\Line(#1,#2)(#3,#4)}
\newcommand{\CrossedCircle}[6]{\BCirc(#1,#2){5}\Line(#3,#6)(#5,#4)\Line(#5,#6)(#3,#4)}
\newcommand{\Higgsino}[5]{\Photon(#1,#2)(#3,#4){3}{#5}\Line(#1,#2)(#3,#4)}
\newcommand{\myGluon}[5]{\Gluon(#1,#2)(#3,#4){3}{#5}}
\newcommand{\Squark}[4]{\DashArrowLine(#1,#2)(#3,#4){2}}
\newcommand{\Squarknoarrow}[4]{\DashLine(#1,#2)(#3,#4){2}}
\newcommand{\Quarknoarrow}[4]{\ArrowLine(#1,#2)(#3,#4)}


\vspace*{-2.5cm}

\begin{flushright}
PSI--PR--17--18 \\
ZU--TH 33/17 \\
FR--PHENO--2017--019 \\
\end{flushright}

\begin{center}
{\large \sc Refinements of the Bottom and Strange MSSM
\\[0.3cm] Higgs Yukawa Couplings at NNLO} \\[1cm]
\end{center}

\begin{center} {\sc Margherita Ghezzi$^1$, Seraina Glaus$^{2,3}$, Dario
M\"uller$^{4}$, Timo Schmidt$^5$ and Michael Spira$^1$} \\[0.8cm]

\begin{small}
{\it \small
$^1$ Institut f\"ur Theoretische Physik, Eberhard Karls
Universit\"at T\"ubingen, D--72076 T\"ubingen, Germany \\
{\it $^2$ Institute for Theoretical Physics, Karlsruhe Institute of
Technology, D--76131 Karlsruhe, Germany}\\
{\it $^3$ Institute for Nuclear Physics, Karlsruhe Institute of
Technology, D--76344 Karlsruhe, Germany} \\
$^4$ Institut f\"ur Theoretische Physik, Zurich University, CH--8057
Zurich, Switzerland \\
$^5$ Albert-Ludwigs-Universit\"at Freiburg, Physikalisches Institut,
D--79104 Freiburg, Germany \\
$^6$ Paul Scherrer Institut, CH--5232 Villigen PSI, Switzerland}
\end{small}
\end{center}


\begin{abstract}
\noindent
We extend the already existing two-loop calculation of the effective
bottom-Yukawa coupling in the MSSM. In addition to the resummation of
the dominant corrections for large values of tg$\beta$, we include the
subleading terms related to the trilinear Higgs coupling $A_b$ and
contributions induced by the electroweak gauge couplings. This
calculation has been extended to the NNLO corrections to the MSSM
strange-Yukawa coupling. Our analysis leads to residual theoretical
uncertainties of the effective Yukawa couplings at the per-cent level.
\end{abstract}



\section{Introduction}
The discovery of a Standard-Model-like Higgs boson at the LHC
\cite{discovery} completed the theory of electroweak and strong
interactions. The existence of an elementary Higgs boson \cite{hi64} is
a necessary ingredient of a weakly interacting renormalizable theory
with spontaneous symmetry breaking \cite{smren}. The measured Higgs mass
of ($125.09 \pm 0.24$) GeV \cite{couplings} ranges at the order of the
weak scale. However, if embedded in a Grand Unified Theory (GUT),
radiative corrections tend to push the Higgs mass towards the GUT scale,
if the Higgs couples to particles at this large scale. This problem is
known as the hierarchy problem \cite{hierarchy}. A solution to this
problem might be offered by supersymmetry (SUSY) at the TeV scale
\cite{susy,susyrev}.

The minimal supersymmetric extension of the Standard Model (MSSM)
contains two Higgs doublets yielding five elementary Higgs bosons after
electroweak symmetry breaking, two neutral CP-even (scalar) bosons $h$,
$H$, one neutral CP-odd (pseudoscalar) boson $A$ and two charged bosons
$H^\pm$. The MSSM Higgs sector is described by two input parameters at
leading order, which are generally chosen as $\tgb=v_2/v_1$, the ratio
of the two vacuum expectation values $v_{1,2}$, and the pseudoscalar
Higgs mass $M_A$, if all SUSY parameters are real. Including the
one-loop and leading two-loop corrections, the upper bound on the light
scalar Higgs mass is lifted to $M_h\lesssim 135$ GeV \cite{mssmrad}. The
more recent three-loop results modify this upper bound by less than 1
GeV \cite{mssmrad3}. The Higgs couplings to gauge bosons and fermions
involve mixing angles $\alpha$ and $\beta$, which are determined by
diagonalizing the neutral and charged Higgs mass matrices. They are
listed in Table \ref{tab:higgs_sm_couplings} relative to the SM Higgs
couplings.%
\begin{table}[t]
\begin{center}
\begin{tabular}{|lc||ccc|} \hline
\multicolumn{2}{|c||}{$\Phi$} & $g^\Phi_u$ & $g^\Phi_d$ &  $g^\Phi_V$ \\
\hline \hline
SM~ & $H$ & 1 & 1 & 1 \\ \hline
MSSM~ & $h$ & $\cos\alpha/\sin\beta$ & $-\sin\alpha/\cos\beta$ & $\sin(\beta-\alpha)$ \\
& $H$ & $\sin\alpha/\sin\beta$ & $\cos\alpha/\cos\beta$ & $\cos(\beta-\alpha)$ \\
& $A$ & $ 1/\tgb$ & $\tgb$ & 0 \\ \hline
\end{tabular}
\end{center}
\vspace*{-0.4cm}
\caption{\it MSSM Higgs couplings to SM particles relative to the SM Higgs
couplings.}
\label{tab:higgs_sm_couplings}
\end{table}
For large $\tgb$ values the down-type Yukawa couplings are enhanced and
the up-type Yukawa couplings suppressed, if the light (heavy) scalar
Higgs mass does not range at its upper (lower) bound, where the
couplings become Standard-Model-like (up to a sign for the heavy scalar
Higgs boson). At present and future colliders this property leads to the
dominance of bottom-Yukawa-coupling induced processes for large $\tgb$
values as Higgs decays into bottom quarks and Higgs bremsstrahlung off
bottom quarks at hadron and $e^+e^-$ colliders. In addition, Higgs-boson
production via gluon fusion $gg\to h,H,A$ is dominantly induced by the
bottom-loop contributions for large $\tgb$. The strongly enhanced
strange-Yukawa coupling on the other hand plays a role for the
subleading charged Higgs decay mode $H^+\to c\bar s$ that can reach
branching ratios at the per-cent level or the reverse charged Higgs
production process $c\bar s\to H^+$ \cite{cs2hc}.

The soft SUSY-breaking terms in the MSSM induce mixing of the current
sfermion eigenstates $\tilde{f}_L$ and $\tilde{f}_R$. The sfermion mass
matrix in the current eigenstate basis is given by\footnote{For
convenience, the $D$-terms have been absorbed in the soft SUSY-breaking
sfermion mass parameters $M_{\tilde f_{L/R}}^2$.}
\beq
{\cal M}^2_{\tilde f}  = \left(\begin{array}{cc}M_{LL}^2 &
M_{LR}^2 \\M_{RL}^2 & M_{RR}^2 \\\end{array} \right) =
\left(\begin{array}{cc} M_{\tilde f_L}^2 + m_f^2\ & m_f(A_f-\mu r_f)
\\ m_f(A_f-\mu r_f)\ & M_{\tilde f_R}^2 + m_f^2 \\\end{array}\right)
\label{eq:MLL_MRR_MRL}
\eeq
with the factors $r_d=1/r_u=\tb$ for down- and up-type sfermions. The
parameter $\mu$ denotes the higgsino mass parameter of the
superpotential. The sfermion mass eigenstates $\tilde f_{1,2}$ emerge
from the current eigenstates $\tilde f_{L,R}$ through rotations by
mixing angles $\theta_f$,
\begin{eqnarray}
\tilde f_1 & = & \tilde f_L \cos\theta_f + \tilde f_R \sin \theta_f
\nonumber \\
\tilde f_2 & = & -\tilde f_L\sin\theta_f + \tilde f_R \cos \theta_f
\label{eq:sfmix}
\end{eqnarray}
which scale with the masses $m_f$ of the SM fermions. Mixing effects are
only relevant for the third-generation sfermions $\tilde t, \tilde b,
\tilde \tau$ and will thus be neglected for the strange squarks in this
work. The mixing angles are determined by
\begin{equation}
\sin 2\theta_f = \frac{2m_f (A_f-\mu r_f)}{m_{\tilde f_1}^2 - m_{\tilde
f_2}^2}
~~~,~~~
\cos 2\theta_f = \frac{M_{\tilde f_L}^2 - M_{\tilde f_R}^2}{m_{\tilde
f_1}^2
- m_{\tilde f_2}^2}
\label{eq:angle}
\end{equation} 
while the masses of the squark mass eigenstates read as
\begin{equation}
m_{\tilde f_{1,2}}^2 = m_f^2 + \frac{1}{2}\left[ M_{\tilde f_L}^2 +
M_{\tilde f_R}^2 \mp \sqrt{(M_{\tilde f_L}^2 - M_{\tilde f_R}^2)^2 +
4m_f^2 (A_f - \mu r_f)^2} \right]
\end{equation}

The topic of this work is the extension of the
next-to-next-to-leading-order (NNLO) SUSY--QCD and top-induced
SUSY--electroweak corrections of the effective bottom-Yukawa couplings
\cite{deltabnnlo} to the terms induced by the soft SUSY-breaking
trilinear coupling $A_b$, the electroweak couplings
$\alpha_1={g'}^2/(4\pi)$, $\alpha_2=g^2/(4\pi)$ [$g,g'$ being the
isospin and hypercharge gauge couplings, respectively] and to the
SUSY--QCD corrections of the strange-Yukawa couplings. The results will
play a role in all processes to which the bottom- and strange-Yukawa
couplings contribute.  In particular the neutral and charged Higgs decay
widths and Higgs radiation off bottom quarks at hadron colliders, which
constitutes the dominant Higgs boson production channel for large $\tb$
at the LHC \cite{review}, are affected.

\section{Effective Bottom- and Strange-Yukawa Couplings}
The dominant parts of the SUSY--QCD (and SUSY--electroweak) corrections
to processes mediated by the bottom- and strange-Yukawa-couplings can be
described in terms of effective bottom- and strange-Yukawa couplings.
These corrections arise in the limit of heavy supersymmetric particles
relative to the energy scale of the particular process. The reliability
of this large mass approximation has been analyzed for neutral MSSM
Higgs decays into bottom quarks $h/H/A\to b\bar b$ \cite{GHS}, charged
Higgs decays to top and bottom quarks $H^\pm\to tb$ \cite{CGNW} and
Higgs radiation off bottom quarks at $e^+e^-$ colliders \cite{ee2hbb}
and hadron colliders \cite{pp2hbb, pp2hbb2} by comparing to the full
next-to-leading-order (NLO) results. For large values of $\tgb$ the
approximation agrees with the NLO results at the sub-per-cent level.

\subsection{Effective Lagrangian}
The dominant contributions to the MSSM bottom- and strange-Yukawa
couplings can be obtained from the effective Lagrangian \cite{GHS,CGNW}
\bea
{\cal L}_{eff} & = & -\sum_{q=b,s} \lambda_q \overline{q_R} \left[
(1+\Delta_{q,1}) \phi_1^0
+ \Delta_{q,2} \phi_2^{0*} \right] q_L + h.c. \nonumber \\
& = & -\sum_{q=b,s} m_q \bar q \left[1+i\gamma_5 \frac{G^0}{v}\right] q
-\frac{m_q/v}{1+\Delta_q} \bar q \left[ g_q^h \left(
1-\frac{\Delta_q}{\tga~\tgb}\right) h \right. \nonumber \\
& & \hspace*{2cm} \left. + g_q^H \left( 1+\Delta_q
\frac{\tga}{\tgb}\right) H
- g_q^A \left(1-\frac{\Delta_q}{\stgb} \right) i \gamma_5 A \right] q
\label{eq:leff}
\eea
with the individual leading one-loop terms for the bottom Yukawa
couplings ($C_F= 4/3$)
\cite{Hall:1993gn}
\begin{eqnarray}
\Delta_{b,1} & = &
-\frac{C_F}{2}~\frac{\alpha_s(\mu_R)}{\pi}~m_{\sgl}~A_b~
I(m^2_{\sbottom_1},m^2_{\sbottom_2},m^2_{\sgl}) \nonumber \\
\Delta_{b,2} & = & \Delta_{b,2}^{QCD} + \Delta_{b,2}^{elw,t} +
\Delta_{b,2}^{elw,1} + \Delta_{b,2}^{elw,2} \nonumber \\
\Delta_{b,2}^{QCD} & = &
\frac{C_F}{2}~\frac{\alpha_s(\mu_R)}{\pi}~m_{\sgl}~\mu~
I(m^2_{\sbottom_1},m^2_{\sbottom_2},m^2_{\sgl}) \nonumber \\
\Delta_{b,2}^{elw,t} & = &
\frac{\lambda_t^2(\mu_R)}{(4\pi)^2}~A_t~\mu~
I(m^2_{\stopx_1},m^2_{\stopx_2},\mu^2) \nonumber \\
\Delta_{b,2}^{elw,1} & = &
-\frac{\alpha_1}{12\pi}~M_1~\mu~\left\{
\frac{1}{3}I(m_{\tilde b_1}^2,m_{\tilde b_2}^2,M_1^2)
+\left( \frac{c_b^2}{2}+s_b^2\right)I(m_{\tilde b_1}^2,M_1^2,\mu^2)
\right. \nonumber \\
& & \left. \hspace*{4cm} +\left( \frac{s_b^2}{2}+c_b^2\right)
I(m_{\tilde b_2}^2,M_1^2,\mu^2)
\right\} \nonumber \\
\Delta_{b,2}^{elw,2} & = &
-\frac{\alpha_2}{4\pi}~M_2~\mu~\left\{
c_t^2 I(m_{\tilde t_1}^2,M_2^2,\mu^2)
+ s_t^2 I(m_{\tilde t_2}^2,M_2^2,\mu^2)
\right. \nonumber \\
& & \left. \hspace*{4cm} +\frac{c_b^2}{2} I(m_{\tilde
b_1}^2,M_2^2,\mu^2)
+\frac{s_b^2}{2} I(m_{\tilde b_2}^2,M_2^2,\mu^2)
\right\}
\label{eq:effparb}
\end{eqnarray}
and for the strange Yukawa couplings\footnote{Due to the tiny charm
Yukawa coupling $\lambda_c$ we neglect electroweak corrections induced
by $\lambda_c$. Moreover, for the strange and charm squarks we neglect
mixing effects, i.e.~work with $c_{s/c}^2 = 1$ and $s_{s/c}^2 = 0$.}
\begin{eqnarray}
\Delta_{s,1} & = &
-\frac{C_F}{2}~\frac{\alpha_s(\mu_R)}{\pi}~m_{\sgl}~A_s~
I(m^2_{\strange_1},m^2_{\strange_2},m^2_{\sgl}) \nonumber \\
\Delta_{s,2} & = & \Delta_{s,2}^{QCD} +
\Delta_{s,2}^{elw,1} + \Delta_{s,2}^{elw,2} \nonumber \\
\Delta_{s,2}^{QCD} & = &
\frac{C_F}{2}~\frac{\alpha_s(\mu_R)}{\pi}~m_{\sgl}~\mu~
I(m^2_{\strange_1},m^2_{\strange_2},m^2_{\sgl}) \nonumber \\
\Delta_{s,2}^{elw,1} & = &
-\frac{\alpha_1}{12\pi}~M_1~\mu~\left\{
\frac{1}{3}I(m_{\tilde s_1}^2,m_{\tilde s_2}^2,M_1^2)
+\left( \frac{c_s^2}{2}+s_s^2\right)I(m_{\tilde s_1}^2,M_1^2,\mu^2)
\right. \nonumber \\
& & \left. \hspace*{4cm} +\left( \frac{s_s^2}{2}+c_s^2\right)
I(m_{\tilde s_2}^2,M_1^2,\mu^2)
\right\} \nonumber \\
\Delta_{s,2}^{elw,2} & = &
-\frac{\alpha_2}{4\pi}~M_2~\mu~\left\{
c_c^2 I(m_{\tilde c_1}^2,M_2^2,\mu^2)
+ s_c^2 I(m_{\tilde c_2}^2,M_2^2,\mu^2)
\right. \nonumber \\
& & \left. \hspace*{4cm} +\frac{c_s^2}{2} I(m_{\tilde
s_1}^2,M_2^2,\mu^2)
+\frac{s_s^2}{2} I(m_{\tilde s_2}^2,M_2^2,\mu^2)
\right\}
\label{eq:effpars}
\end{eqnarray}
where $s_q = \sin\theta_q$, $c_q = \cos\theta_q$ ($q=t,b,c,s$) are
related to the squark mixing angles $\theta_q$ of Eq.~(\ref{eq:angle}).
The final contribution in the mass-eigenstate-basis can be derived as
\bea
\Delta_q = \frac{\Delta_{q,2}~\tgb}{1+\Delta_{q,1}} \qquad\qquad (q=b,s)
\label{eq:deltab0}
\eea
The auxiliary function $I$ is given by
\beq
I(a,b,c) = \frac{\displaystyle ab\log\frac{a}{b} + bc\log\frac{b}{c}
+ ca\log\frac{c}{a}}{(a-b)(b-c)(a-c)}
\eeq
The field amplitudes $\phi_1^0$ and $\phi_2^0$ of the neutral
components of the Higgs doublets that couple to down- and up-type
quarks, respectively, are transformed to the mass eigenstates
$h,H,A$ by the mixing angles $\alpha,\beta$
\bea
\phi_1^0 & = & \frac{1}{\sqrt{2}}\left[ v_1 + H\coa - h\sia + iA\sib -
iG^0\cob
\right] \nonumber \\
\phi_2^0 & = & \frac{1}{\sqrt{2}}\left[ v_2 + H\sia + h\coa + iA\cob +
iG^0\sib
\right]
\eea
The two vacuum expectation values are connected to the Fermi constant
$G_F$ by $v^2={v_1^2+v_2^2} = 1/({\sqrt{2}G_F})$. The would-be Goldstone
field $G^0$ is 'eaten' by the $Z$ boson and builds up its longitudinal
component. The top-Yukawa coupling $\lambda_t$ determines the top mass
by $m_t = \lambda_t v_2/\sqrt{2}$ at leading order. The soft
SUSY-breaking trilinear couplings of the top, bottom and strange squarks
are denoted by $A_t, A_b$ and $A_s$, the higgsino mass parameter by
$\mu$ and the strong coupling constant by $\alpha_s$. The
renormalization scale is depicted as $\mu_R$. The corrections
$\Delta_{b,s}$ modify the relation between the bottom (strange) quark
mass $m_b~(m_s)$ and the bottom (strange) Yukawa coupling
$\lambda_b~(\lambda_s)$,
\beq
m_q = \frac{\lambda_q v_1}{\sqrt{2}} \left[ 1 + \Delta_{q,1} +
\Delta_{q,2}\,\tgb \right] \qquad\qquad (q=b,s)
\eeq
The effective Lagrangian of Eq.~(\ref{eq:leff}) can be expressed as
(omitting the mass and Goldstone terms)
\beq
{\cal L}_{eff} = -\sum_{q=b,s} \frac{m_q}{v}\ \bar{q}\ \big[\ \tilde
g^h_q\ h + \tilde g^H_q\ H - \tilde g^A_q\ i\gamma_5\ A\ ]\ q
\label{eq:leff2}
\eeq
with the effective (resummed) couplings
\bea
\tilde g^h_q & = & \frac{g^h_q}{1+\Delta_q}\left[ 1 -
\frac{\Delta_q}{\tga\tgb}  \right] \nonumber \\
\tilde g^H_q & = & \frac{g^H_q}{1+\Delta_q}\left[ 1 + \Delta_q
\frac{\tga}{\tgb} \right] \nonumber \\
\tilde g^A_q & = & \frac{g^A_q}{1+\Delta_q}\left[ 1 -
\frac{\Delta_q}{\tgb^2} \right]
\label{eq:rescoup}
\eea
Even though the SUSY corrections $\Delta_q$ are loop-suppressed, they
are significant for large values of $\tgb$.  In these regions they
dominate the supersymmetric corrections to the bottom- and
strange-Yukawa couplings. The effective Lagrangian in
Eq.~(\ref{eq:leff}) has been derived by integrating out the heavy SUSY
particles so that it is not only valid for large values of $\tgb$.  By
using power counting it has been shown that the Lagrangian of
Eq.~(\ref{eq:leff}) resums all terms of ${\cal
O}\left[(\alpha_s\,\mu\,\tgb)^n\right]$ and ${\cal
O}\left[(\alpha_s\,A_{b,s})^n\right]$. For the bottom-Yukawa coupling a
resummation of the ${\cal O}\left[(\lambda^2_t\,A_t\,\tgb)^n\right]$
terms is achieved in addition \cite{CGNW, GHS} (including mixed
contributions).

\subsection{Low Energy Theorems} \label{sc:let}
The derivation of higher-order corrections to the effective Yukawa
couplings would need the calculation of the related three-point
functions in the low-energy limit. This, however, can be reduced to the
determination of self-energy diagrams by means of low energy theorems
\cite{let}. These are based on the feature that in the limit of
vanishing Higgs momentum, matrix elements with an external Higgs boson
can be generated from the corresponding matrix elements without the
external Higgs particle by the shifts $v_1\to \sqrt{2}\phi_1^0$ and
$v_2\to \sqrt{2}\phi_2^{0*}$. Thus, only the calculation of the related
parts of the bottom and strange quark self-energies is required. The
dominant parts $\Delta_{q,1/2}~(q=b,s)$ originate from the scalar part
$\Sigma_S(m_q^2)$ of the self-energy\footnote{The fermionic self-energy
can be split into a scalar, vectorial and axial-vectorial part as
$\Sigma(p) = \Sigma_S(p^2) + \slashed{p}\,\Sigma_V(p^2) +
\slashed{p}\gamma_5\,\Sigma_A(p^2)$.}. This affects the relation between
the bottom (strange) Yukawa coupling $\lambda_b~(\lambda_s)$ and the
mass $m_q$ of the bottom (strange) quark,
\beq
m_q=\frac{\lambda_q}{\sqrt{2}}v_1 + \Sigma_S(m_q^2)
\label{eq:sigma_s}
\eeq
with the dominant terms of the self-energy $\Sigma_S(m_q^2)$ for heavy
SUSY particles
\bea
\Sigma_S(m_q^2) & = & \frac{\lambda_q}{\sqrt{2}}\,v_1\ \left[
\Delta_{q,1} + \Delta_{q,2} \tgb \right]
\eea
The NLO--QCD parts of $\Delta_b$ and $\Delta_s$ in Eq.~(\ref{eq:leff})
can be obtained from off-diagonal mass insertions of the type $\lambda_q
(A_q v_1 - \mu v_2)$ (up to a factor $1/\sqrt{2}$) in the squark
propagators, as shown in Fig.~\ref{fig:1loop_Selfenergies} at one-loop
order.%
\begin{figure}[t]
\begin{center}
\begin{picture}(300,60)(0,0)
\Text(0,40)[lb]{(a)}
\Quark{0}{0}{40}{0}
\Gluino{40}{0}{100}{0}{6}
\Quark{100}{0}{140}{0}
\DashCArc(70,0)(30,0,180){2}
\CrossedCircle{70}{30}{67}{27}{73}{33}
\Text(20,3)[cb]{$\bL$}
\Text(120,3)[cb]{$\bR$}
\Text(45,20)[rb]{$\bsL$}
\Text(95,20)[lb]{$\bsR$}
\Text(70,-5)[ct]{$\gluino$}
\Text(70,40)[cb]{$\lb (A_b v_1 - \mu v_2)$}
\Text(160,40)[lb]{(b)}
\Quark{160}{0}{200}{0}
\Gluino{200}{0}{260}{0}{6}
\Quark{260}{0}{300}{0}
\DashCArc(230,0)(30,0,180){2}
\CrossedCircle{230}{30}{227}{27}{233}{33}
\Text(180,3)[cb]{$\sL$}
\Text(280,3)[cb]{$\sR$}
\Text(205,20)[rb]{$\strL$}
\Text(255,20)[lb]{$\strR$}
\Text(230,-5)[ct]{$\gluino$}
\Text(230,40)[cb]{$\ls (A_s v_1 - \mu v_2)$}
\end{picture}
\end{center}
\caption{\it One-loop diagrams of the SUSY--QCD contributions to (a) the
bottom and (b) the strange self-energies with the off-diagonal mass
insertions related to the $\Delta_q~(q=b,s)$ corrections of the
bottom- and strange-Yukawa couplings. The contributing particles involve
bottom and strange quarks $b,s$ and squarks $\sbottom,\strange$ as well
as gluinos $\gluino$.}
\label{fig:1loop_Selfenergies}
\end{figure}
This results in the finite expressions of Eqs.~(\ref{eq:effparb},
\ref{eq:effpars}) (supplemented by the SUSY--electroweak corrections
originating from higgsino, wino and bino exchange to the bottom-Yukawa
couplings) after transforming the fields from current-eigenstates to the
mass eigenstates. These contributions are not renormalized at NLO due to
the absence of tree-level bottom and strange couplings involving $A_q$
or $\mu$.

\section{NNLO Corrections}
The NNLO analysis of the effective bottom- and strange-Yukawa
couplings necessitates the calculation of the leading NNLO corrections to
the bottom and strange self-energies. The NNLO expressions of the
bottom-Yukawa couplings have been obtained in
Refs.~\cite{deltabnnlo,deltabnnlo2}. We will extend these results to the
non-$\tgb$-enhanced $A_b$ terms and to the strange Yukawa couplings. In
this work we will neglect intergenerational mixing so that issues
related to the flavour sector \cite{mfv} can be disregarded.

\subsection{Bottom-Yukawa Couplings}
\subsubsection{SUSY--QCD Corrections to $\Delta_{b,1}$}
Since the effective insertions according to
Fig.~\ref{fig:1loop_Selfenergies} are always proportional to $A_b -
\mu\,\tgb$, the contributions of all two-loop diagrams for the
bottom-Yukawa coupling is the same for the $A_b$ and the $\mu\,\tgb$
contributions (up to the relative overall sign). The renormalization
proceeds along the lines of Ref.~\cite{deltabnnlo} so that the SUSY--QCD
corrections to the $\Delta_{b,1}$ terms are the same as for the
$\Delta_{b,2}$ contributions after renormalization (including the
SUSY-restoring counter terms \cite{susyrest}). Denoting the
(renormalized) NNLO-corrected SUSY--QCD part $\Delta_{b,2}^{QCD}$ of
Ref.~\cite{deltabnnlo} as
\bea
\Delta_{b,2}^{QCD} = \mu\,\Delta^{NLO} \left[ 1+\delta_b \right]
\eea
with the NNLO correction $\delta_b$ and
\bea
\Delta^{NLO} = \frac{C_F}{2}~\frac{\alpha_s(\mu_R)}{\pi}~m_{\sgl}~
I(m^2_{\sbottom_1},m^2_{\sbottom_2},m^2_{\sgl})
\eea
the effective correction to the bottom-Yukawa couplings of
Eq.~(\ref{eq:deltab0}) acquires the form
\bea
\Delta_b = \frac{\mu~\Delta^{NLO} \left[ 1+\delta_b \right] +
\Delta_{b,2}^{elw,t} \left[ 1+\delta_t \right] + \Delta_{b,2}^{elw,1}
\left[ 1+\delta_1 \right] + \Delta_{b,2}^{elw,2} \left[ 1+\delta_2
\right]}{1-A_b^0~\Delta^{NLO} \left[ 1+\delta_b \right]}~\tgb
\label{eq:deltab}
\eea
where $A_b^0$ denotes the bare trilinear coupling that is renormalized
in SUSY--QCD and $\delta_t$ the SUSY--QCD corrections to
$\Delta_{b,2}^{elw,t}$ \cite{deltabnnlo}. The NNLO SUSY--QCD corrections
$\delta_1$ ($\delta_2$) to $\Delta_{b,2}^{elw,1}$
($\Delta_{b,2}^{elw,2}$) will be derived and discussed in the next
subsection. The renormalization of $A_b$ emerges from a non-leading
order contribution in our context: for the $\overline{\rm
MS}$-renormalized trilinear coupling within dimensional regularization
in $n=4-2\epsilon$ dimensions we obtain
\bea
A_b^0 & = & A_b(\mu_R^2) + \delta A_b \nonumber \\
\delta A_b & = & C_F \frac{\alpha_s}{\pi} \Gamma(1+\epsilon) \left(
\frac{4\pi \mu^2}{\mu_R^2} \right)^\epsilon \frac{m_{\sgl}}{\epsilon}
\neq {\cal O} (A_b)
\eea
so that $A_b$ is not renormalized at ${\cal O}(\alpha_s\,A_b)$. We have
explicitly checked that the divergence corresponding to the counter
term of $A_b$ is generated by the diagram of Fig.~\ref{fg:abdia} with
an insertion $\lambda_b\,v_1$ (up to a factor $1/\sqrt{2}$) at the
virtual bottom-quark line. Thus the final expression including the
${\cal O}(A_b)$ terms is given by Eq.~(\ref{eq:deltab}) with $A_b^0$
replaced by the renormalized $A_b(\mu_R^2)$ coupling,
\bea
\Delta_b = \frac{\Delta_{b,2}^{QCD} \left[ 1+\delta_b \right] +
\Delta_{b,2}^{elw,t} \left[ 1+\delta_t \right] + \Delta_{b,2}^{elw,1}
\left[ 1+\delta_1 \right] + \Delta_{b,2}^{elw,2} \left[
1+\delta_2 \right]} {1+\Delta_{b,1} \left[ 1+\delta_b
\right]}~\tgb
\eea
with $\Delta_{b,1}, \Delta_{b,2}^{QCD}, \Delta_{b,2}^{elw,t},
\Delta_{b,2}^{elw,1}$ and $\Delta_{b,2}^{elw,2}$ defined in
Eq.~(\ref{eq:effparb}).
\begin{figure}[htbp]
\begin{picture}(200,90)(-150,0)
\ArrowLine(0,0)(40,0)
\DashLine(40,0)(70,0){5}
\DashLine(130,0)(160,0){5}
\ArrowLine(70,0)(100,0)
\ArrowLine(100,0)(130,0)
\CArc(100,0)(30,0,180)
\GlueArc(100,0)(30,0,180){5}{7}
\DashLine(130,0)(150,0){5}
\ArrowLine(160,0)(200,0)
\CArc(100,0)(60,0,180)
\GlueArc(100,0)(60,0,180){5}{15}
\CrossedCircle{100}{0}{97}{-3}{103}{3}
\put(10,6){$b_L$}
\put(185,6){$b_R$}
\put(50,-14){$\tilde b_L$}
\put(142,-14){$\tilde b_R$}
\put(80,6){$b_L$}
\put(110,6){$b_R$}
\put(90,-15){$\lambda_b v_1$}
\put(94,40){$\tilde g$}
\put(55,55){$\tilde g$}
\end{picture} \\
\caption{\it Two-loop diagram of sbottom-self-energy insertions
contributing to the SUSY--QCD corrections to the bottom-quark
self-energy. This involves bottom quarks $b$, bottom squarks $\sbottom$ and
gluinos $\sgl$.}
\label{fg:abdia}
\end{figure}

In this work we adopt the renormalization program of Ref.~\cite{hsqsq},
i.e.~the counter term for the top-Yukawa-induced electroweak
contributions $\Delta_{b,2}^{elw,t}$ is modified for the trilinear
coupling $A_t$ that is defined in the $\overline{\rm MS}$ scheme leading
to a vanishing counter term for $A_t$ at the order we are calculating.

\subsubsection{SUSY--QCD Corrections to $\Delta^{elw,1/2}_{b,2}$}
For the calculation of the SUSY--QCD corrections to the terms
$\Delta_{b,2}^{elw,1}$ and $\Delta_{b,2}^{elw,1}$ we will reduce the
associated higgsino propagators to the contributions relevant for the
$\tgb$-enhanced corrections. For the neutralinos in the basis $(\tilde
B, \tilde W^3, \tilde H_1^0, \tilde H_2^0)$, the full inverse
propagator matrix is given by
\bea
{\cal P}^{-1} & = & {\cal P}_0^{-1} + {\cal D} \nonumber \\[0.5cm]
{\cal P}_0^{-1} & = & \left( \begin{array}{cccc}
\not\! p-M_1 & 0 & 0 & 0 \\
0 & \not\! p-M_2 & 0 & 0 \\
0 & 0 & \not\! p & \mu \\
0 & 0 & \mu & \not\! p
\end{array} \right) \nonumber \\[0.5cm]
{\cal D} & = & \left( \begin{array}{cccc} \displaystyle
0 & 0 & \displaystyle \frac{g' v}{2} c_\beta & \displaystyle -\frac{g'
v}{2} s_\beta \\[0.5cm]
0 & 0 & \displaystyle -\frac{g v}{2} c_\beta & \displaystyle \frac{g
v}{2} s_\beta \\[0.5cm]
\displaystyle \frac{g' v}{2} c_\beta & \displaystyle -\frac{g v}{2}
c_\beta & 0 & 0 \\[0.5cm]
\displaystyle -\frac{g' v}{2} s_\beta & \displaystyle \frac{g v}{2} s_\beta & 0 & 0
\end{array} \right)
\eea
where the part ${\cal D}$ is subleading in the limit of heavy SUSY
particles that we are working in. Keeping only linear terms in ${\cal
D}$, the propagator matrix is then given by
\bea
{\cal P} & = & {\cal P}_0 - {\cal P}_0 {\cal D}{\cal P}_0  \nonumber \\[0.5cm]
{\cal P}_0 & = & \left( \begin{array}{cccc}
\displaystyle \frac{\not\! p+M_1}{p^2-M_1^2} & 0 & 0 & 0 \\[0.5cm]
0 & \displaystyle \frac{\not\! p+M_2}{p^2-M_2^2} & 0 & 0 \\[0.5cm]
0 & 0 & \displaystyle \frac{\not\! p}{p^2-\mu^2} & \displaystyle
-\frac{\mu}{p^2-\mu^2} \\[0.5cm]
0 & 0 & \displaystyle -\frac{\mu}{p^2-\mu^2} & \displaystyle \frac{\not\!
p}{p^2-\mu^2}
\end{array} \right) \nonumber \\[0.5cm]
-{\cal P}_0 {\cal D}{\cal P}_0 & = & \left( \begin{array}{cc} \displaystyle
0 & {\cal A} \\ {\cal A}^T & 0 \end{array} \right) \nonumber \\[0.5cm]
{\cal A} & = & \frac{c_\beta}{p^2-\mu^2} \left(
\begin{array}{cc} \displaystyle
\displaystyle -\frac{g' v}{2}~\frac{(\not\! p+M_1)(\not\! p+\mu\tgb)}{p^2-M_1^2} &
\displaystyle \frac{g' v}{2}~\frac{(\not\! p+M_1)(\not\! p
\tgb+\mu)}{p^2-M_1^2} \\[0.5cm]
\displaystyle \frac{g v}{2}~\frac{(\not\! p+M_2)(\not\! p+\mu\tgb)}{p^2-M_2^2} &
\displaystyle -\frac{g v}{2}~\frac{(\not\! p+M_2)(\not\! p
\tgb+\mu)}{p^2-M_2^2}
\end{array} \right) \, .
\label{eq:propn}
\eea
Inserting the corresponding diagonal and off-diagonal propagators into
the self-energies of Fig.~\ref{fg:diaal1} and keeping only the
$\tgb$-enhanced terms in the numerators of the off-diagonal propagators
contained in ${\cal A}$, we arrive at the proper expressions for
$\Delta_{b,2}^{elw,1}$ of Eq.~(\ref{eq:effparb}) where the identities
\bea
M_{LL}^2 & = & M_{\squark_L}^2 + m_q^2 = m_{\squark_1}^2
c_q^2 + m_{\squark_2}^2 s_q^2 \nonumber \\
M_{RR}^2 & = & M_{\squark_R}^2 + m_q^2 = m_{\squark_1}^2
s_q^2 + m_{\squark_2}^2 c_q^2 \qquad (q=b,t)
\label{eq:mllrr}
\eea
have been used. Including only the $\tgb$-enhanced contributions ensures that
terms proportional to $v_2$ are kept thanks to the relation $v\,c_\beta
\tgb = v_2$ that are then shifted by the full Higgs field $v_2 \to
\sqrt{2}\phi_2^{0*}$ according to the discussion of Section \ref{sc:let}.
\begin{figure}[htbp]
\begin{picture}(150,40)(10,0)
\SetScale{0.8}
\ArrowLine(20,0)(50,0)
\Line(50,0)(150,0)
\Photon(50,0)(150,0){3}{10}
\DashCArc(100,0)(50,0,180){5}
\ArrowLine(150,0)(180,0)
\CrossedCircle{100}{50}{97}{47}{103}{53}
\put(24,5){$b_L$}
\put(130,5){$b_R$}
\put(56,20){$\tilde b_L$}
\put(96,20){$\tilde b_R$}
\put(65,48){$-\lambda_b v_2$}
\put(72,-16){$\tilde B$}
\end{picture}
\begin{picture}(150,40)(0,0)
\SetScale{0.8}
\ArrowLine(20,0)(50,0)
\Line(50,0)(150,0)
\Photon(50,0)(150,0){3}{10}
\DashCArc(100,0)(50,0,180){5}
\ArrowLine(150,0)(180,0)
\CrossedCircle{100}{0}{97}{-3}{103}{3}
\put(24,5){$b_L$}
\put(130,5){$b_R$}
\put(72,24){$\tilde b_L$}
\put(56,-16){$\tilde B$}
\put(96,-16){$\tilde H_1^0$}
\end{picture}
\begin{picture}(150,90)(-10,0)
\SetScale{0.8}
\ArrowLine(20,0)(50,0)
\Line(50,0)(150,0)
\Photon(50,0)(150,0){3}{10}
\DashCArc(100,0)(50,0,180){5}
\ArrowLine(150,0)(180,0)
\CrossedCircle{100}{0}{97}{-3}{103}{3}
\put(24,5){$b_L$}
\put(130,5){$b_R$}
\put(72,24){$\tilde b_R$}
\put(56,-16){$\tilde H_1^0$}
\put(96,-16){$\tilde B$}
\end{picture} \\
\caption{\it One-loop diagrams of sbottom-self-energy insertions
contributing to the SUSY--QCD corrections $\Delta_{b,2}^{elw,1}$ of the
bottom-quark self-energy involving bottom quarks $b$, bottom squarks
$\sbottom$, binos $\tilde B$ and higgsinos $\tilde H_1^0$. The crossed
lines indicate off-diagonal propagator contributions.}
\label{fg:diaal1}
\end{figure}

For the calculation of $\Delta_{b,2}^{elw,2}$ we have to use the
off-diagonal propagators of Eq.~(\ref{eq:propn}) proportional to the
isospin gauge coupling $g$, but have to include chargino propagators for
the diagrams involving top squarks in addition. In the basis $(\tilde
W^+, \tilde H_1^+, \tilde H_2^+)$, the full inverse chargino propagator
matrix is given by
\bea
{\cal P}^{-1} & = & {\cal P}_0^{-1} + {\cal D} \nonumber \\[0.5cm]
{\cal P}_0^{-1} & = & \left( \begin{array}{ccc}
\not\! p-M_2 & 0 & 0 \\
0 & \not\! p & -\mu \\
0 & -\mu & \not\! p
\end{array} \right)\, , \qquad
{\cal D} = \left( \begin{array}{ccc} \displaystyle
0 & \displaystyle -\frac{g v}{\sqrt{2}} c_\beta & \displaystyle -\frac{g
v}{\sqrt{2}} s_\beta \\[0.5cm]
\displaystyle -\frac{g v}{\sqrt{2}} c_\beta & 0 & 0 \\[0.5cm]
\displaystyle -\frac{g v}{\sqrt{2}} s_\beta & 0 & 0
\end{array} \right)
\eea
where the term ${\cal D}$ is subleading in the heavy-SUSY-particle
limit. The propagator matrix is then obtained as
\bea
{\cal P} & = & {\cal P}_0 - {\cal P}_0 {\cal D}{\cal P}_0  \nonumber \\[0.5cm]
{\cal P}_0 & = & \left( \begin{array}{ccc}
\displaystyle \frac{\not\! p+M_2}{p^2-M_2^2} & 0 & 0 \\[0.5cm]
0 & \displaystyle \frac{\not\! p}{p^2-\mu^2} & \displaystyle
\frac{\mu}{p^2-\mu^2} \\[0.5cm]
0 & \displaystyle \frac{\mu}{p^2-\mu^2} & \displaystyle \frac{\not\!
p}{p^2-\mu^2}
\end{array} \right) \nonumber \\[0.5cm]
-{\cal P}_0 {\cal D}{\cal P}_0 & = & {\cal N} \left(
\begin{array}{ccc} \displaystyle
0 & \displaystyle (\not\! p+M_2)(\not\! p+\mu\tgb) &
\displaystyle (\not\! p+M_2)(\not\! p \tgb+\mu) \\[0.5cm]
\displaystyle (\not\! p+\mu\tgb) (\not\! p + M_2) & 0 & 0 \\[0.5cm]
\displaystyle (\not\! p\tgb +\mu) (\not\! p + M_2) & 0 & 0
\end{array} \right)
\eea
where the normalization factor reads
\bea
{\cal N} = \frac{g v c_\beta}{\sqrt{2} (p^2-M_2^2) (p^2-\mu^2)}
\eea
Inserting the corresponding $\tgb$-enhanced terms of the off-diagonal
propagators into the self-energies of Fig.~\ref{fg:diaal2}, we obtain
the proper expressions for $\Delta_{b,2}^{elw,2}$ of
Eq.~(\ref{eq:effparb}).
\begin{figure}[htbp]
\begin{picture}(150,40)(-90,0)
\SetScale{0.8}
\ArrowLine(20,0)(50,0)
\Line(50,0)(150,0)
\Photon(50,0)(150,0){3}{10}
\DashCArc(100,0)(50,0,180){5}
\ArrowLine(150,0)(180,0)
\CrossedCircle{100}{0}{97}{-3}{103}{3}
\put(24,5){$b_L$}
\put(130,5){$b_R$}
\put(72,24){$\tilde b_L$}
\put(56,-16){$\tilde W^3$}
\put(96,-16){$\tilde H_1^0$}
\end{picture}
\begin{picture}(150,40)(-100,0)
\SetScale{0.8}
\ArrowLine(20,0)(50,0)
\Line(50,0)(150,0)
\Photon(50,0)(150,0){3}{10}
\DashCArc(100,0)(50,0,180){5}
\ArrowLine(150,0)(180,0)
\CrossedCircle{100}{0}{97}{-3}{103}{3}
\put(24,5){$b_L$}
\put(130,5){$b_R$}
\put(72,24){$\tilde t_L$}
\put(56,-16){$\tilde W^-$}
\put(96,-16){$\tilde H_1^-$}
\end{picture} \\
\caption{\it One-loop diagrams of sbottom-self-energy insertions
contributing to the SUSY--QCD corrections $\Delta_{b,2}^{elw,2}$ of the
bottom-quark self-energy involving bottom quarks $b$, sbottoms
$\sbottom$, stops $\stopx$, winos $\tilde W^{3,\pm}$ and higgsinos
$\tilde H_1^{0,\pm}$. The crossed lines indicate the $\tgb$-enhanced
off-diagonal wino-higgsino propagator contributions.}
\label{fg:diaal2}
\end{figure}

The leading $\tgb$-enhanced terms of the off-diagonal chargino- and
neutralino-propagator-matrix entries have been used for the calculation
of the two-loop SUSY--QCD corrections of corrections
$\Delta_{b,2}^{elw,1/2}$ to the bottom Yukawa couplings in
Eq.~(\ref{eq:effparb}). The relevant two-loop diagrams of the bottom
self-energy are depicted in Fig.~\ref{fg:2-loop}. The technical method
of our calculation follows the analysis of Refs.~\cite{deltabnnlo}. The
diagrams of Fig.~\ref{fg:2-loop} are evaluated with single crosses in
each sbottom/stop or gaugino/higgsino line individually. However, for
the additional contributions going with $s_b^2, c_b^2, s_t^2, c_t^2$ in
Eq.~(\ref{eq:effparb}) diagrams with three crosses have to be taken into
account, too, as exemplified for one diagram involving top quarks and
stops in Fig.~\ref{fg:crosses}. This means that there are contributions
with three off-diagonal propagators inserted in the two-loop diagrams.
These diagrams complete the corrections to the terms proportional to
$s_b^2, c_b^2, s_t^2, c_t^2$ in Eq.~(\ref{eq:effparb}) by means of the
relations of Eq.~(\ref{eq:mllrr}) and
\bea
M_{LR}^2 & = & m_q (A_q - \mu r_q) = (m_{\squark_1}^2 -
m_{\squark_2}^2) s_q c_q \qquad (q=b,t)
\eea
For the contributions to $\Delta_{b,2}^{elw,2}$, crosses involving the
off-diagonal entry of the top-quark propagator matrix
\bea
{\cal P} = \frac{1}{p^2-m_t^2} \left( \begin{array}{cc}
\not\! p & m_t \\
m_t & \not\! p
\end{array} \right)
\eea
in the $(t_L, t_R)$ basis induced by the large top mass have to be
considered, too. It should be noted that the symmetry factors between
the self-energies and the effective coupling of the Lagrangian have to
be taken into account properly for the shift $v_2\to \sqrt{2}\phi_2^{0*}$.%
\begin{figure}[htbp]
\SetScale{0.8}
\begin{picture}(200,40)(-20,0)
\ArrowLine(0,0)(50,0)
\Line(50,0)(100,0)
\Photon(50,0)(100,0){5}{3}
\ArrowLine(100,0)(150,0)
\DashLine(100,50)(100,0){5}
\ArrowLine(150,0)(200,0)
\GlueArc(100,0)(50,0,90){5}{5}
\DashCArc(100,0)(50,90,180){5}
\put(10,6){$b$}
\put(148,6){$b$}
\put(84,16){$\sbottom / \stopx$}
\put(37,-18){$\tilde B/\tilde W/\tilde H_1$}
\put(100,-12){$b$}
\put(40,32){$\sbottom / \stopx$}
\put(116,32){$g$}
\end{picture}
\begin{picture}(200,40)(-10,0)
\ArrowLine(0,0)(50,0)
\ArrowLine(150,0)(200,0)
\ArrowLine(50,0)(70,0)
\ArrowLine(130,0)(150,0)
\Line(70,0)(130,0)
\Photon(70,0)(130,0){5}{4}
\GlueArc(100,0)(50,0,180){5}{10}
\DashCArc(100,0)(30,0,180){5}
\put(10,6){$b$}
\put(148,6){$b$}
\put(48,-12){$b$}
\put(56,-18){$\tilde B/\tilde W/\tilde H_1$}
\put(112,-12){$b$}
\put(72,10){$\sbottom / \stopx$}
\put(116,34){$g$}
\end{picture} \\
\begin{picture}(200,90)(-20,0)
\ArrowLine(0,0)(50,0)
\Line(50,0)(100,0)
\Photon(50,0)(100,0){5}{3}
\DashLine(100,0)(150,0){5}
\ArrowLine(100,50)(100,0)
\ArrowLine(150,0)(200,0)
\CArc(100,0)(50,0,90)
\GlueArc(100,0)(50,0,90){5}{5}
\DashCArc(100,0)(50,90,180){5}
\put(10,6){$b$}
\put(148,6){$b$}
\put(84,16){$b/t$}
\put(37,-18){$\tilde B/\tilde W/\tilde H_1$}
\put(100,-15){$\sbottom$}
\put(40,32){$\sbottom / \stopx$}
\put(116,32){$\sgl$}
\end{picture}
\begin{picture}(200,90)(-20,0)
\ArrowLine(0,0)(50,0)
\Line(50,0)(150,0)
\Photon(50,0)(150,0){5}{7}
\ArrowLine(150,0)(200,0)
\DashCArc(100,0)(50,0,180){5}
\GOval(100,46)(10,20)(0){0.5}
\put(10,6){$b$}
\put(148,6){$b$}
\put(58,-18){$\tilde B/\tilde W/\tilde H_1$}
\put(33,25){$\sbottom / \stopx$}
\put(116,25){$\sbottom / \stopx$}
\end{picture} \\
\begin{picture}(200,70)(-10,0)
\DashLine(0,0)(100,0){5}
\GOval(50,0)(10,20)(0){0.5}
\DashLine(130,0)(230,0){5}
\GlueArc(180,0)(20,0,180){4}{5}
\DashLine(260,0)(360,0){5}
\ArrowLine(290,0)(330,0)
\CArc(310,0)(20,0,180)
\GlueArc(310,0)(20,0,180){4}{5}
\DashLine(390,0)(490,0){5}
\DashCArc(440,20)(20,0,360){5}
\Vertex(440,0){2}
\put(88,-3){$=$}
\put(193,-3){$+$}
\put(297,-3){$+$}
\put(10,6){$\squark$}
\put(70,6){$\squark$}
\put(113,6){$\squark$}
\put(143,-13){$\squark$}
\put(143,25){$g$}
\put(175,6){$\squark$}
\put(218,6){$\squark$}
\put(248,-12){$q$}
\put(248,25){$\sgl$}
\put(280,6){$\squark$}
\put(323,6){$\squark$}
\put(351,35){$\squark$}
\put(383,6){$\squark$}
\put(420,-2){$(\squark = \sbottom, \stopx)$}
\end{picture} \\
\caption{\it Generic two-loop diagrams of the SUSY--QCD contributions to
the bottom self-energy involving bottom quarks $b$, sbottoms $\sbottom$
and stops $\st$, gluons $g$, gluinos $\sgl$, binos $\tilde B$,
winos $\tilde W$ and higgsinos $\tilde H_1^{0,\pm}$.}
\label{fg:2-loop}
\end{figure}
\begin{figure}[hbt]
\SetScale{0.8}
\begin{picture}(200,40)(-157,0)
\ArrowLine(0,0)(50,0)
\Line(50,0)(100,0)
\Photon(50,0)(100,0){5}{3}
\DashLine(100,0)(150,0){5}
\ArrowLine(100,50)(100,0)
\ArrowLine(150,0)(200,0)
\CArc(100,0)(50,0,90)
\GlueArc(100,0)(50,0,90){5}{5}
\DashCArc(100,0)(50,90,180){5}
\Line(100,-40)(100,-60)
\ArrowLine(100,-60)(100,-61)
\put(10,6){$b$}
\put(148,6){$b$}
\put(84,16){$t$}
\put(45,-18){$\tilde W/\tilde H_1$}
\put(100,-15){$\sbottom$}
\put(40,32){$\stopx$}
\put(116,32){$\sgl$}
\end{picture} \\
\begin{picture}(200,110)(-60,0)
\ArrowLine(0,0)(50,0)
\Line(50,0)(100,0)
\Photon(50,0)(100,0){5}{3}
\DashLine(100,0)(150,0){5}
\ArrowLine(100,50)(100,0)
\ArrowLine(150,0)(200,0)
\CArc(100,0)(50,0,90)
\GlueArc(100,0)(50,0,90){5}{5}
\DashCArc(100,0)(50,90,180){5}
\put(2,6){$b_L$}
\put(148,6){$b_R$}
\put(84,16){$t$}
\put(37,-18){$\tilde W^-$}
\put(67,-18){$\tilde H_1^-$}
\put(100,-15){$\sbottom_R$}
\put(40,32){$\stopx_L$}
\put(116,32){$\sgl$}
\CrossedCircle{75}{0}{72}{-3}{78}{3}
\put(173,-3){+}
\end{picture}
\begin{picture}(200,110)(-50,0)
\ArrowLine(0,0)(50,0)
\Line(50,0)(100,0)
\Photon(50,0)(100,0){5}{3}
\DashLine(100,0)(150,0){5}
\ArrowLine(100,50)(100,25)
\ArrowLine(100,25)(100,0)
\ArrowLine(150,0)(200,0)
\CArc(100,0)(50,0,90)
\GlueArc(100,0)(50,0,90){5}{5}
\DashCArc(100,0)(50,90,180){5}
\put(2,6){$b_L$}
\put(148,6){$b_R$}
\put(86,16){$t$}
\put(37,-18){$\tilde W^-$}
\put(67,-18){$\tilde H_1^-$}
\put(100,-15){$\sbottom_R$}
\put(30,20){$\stopx_L$}
\put(55,43){$\stopx_R$}
\put(116,32){$\sgl$}
\CrossedCircle{65}{35}{62}{32}{68}{38}
\CrossedCircle{75}{0}{72}{-3}{78}{3}
\CrossedCircle{100}{25}{97}{22}{103}{28}
\end{picture} \\
\begin{picture}(200,110)(-60,0)
\ArrowLine(0,0)(50,0)
\Line(50,0)(100,0)
\Photon(50,0)(100,0){5}{3}
\DashLine(100,0)(150,0){5}
\ArrowLine(100,50)(100,0)
\ArrowLine(150,0)(200,0)
\CArc(100,0)(50,0,90)
\GlueArc(100,0)(50,0,90){5}{5}
\DashCArc(100,0)(50,90,180){5}
\put(2,6){$b_R$}
\put(148,6){$b_L$}
\put(84,16){$t$}
\put(37,-18){$\tilde H_1^-$}
\put(67,-18){$\tilde W^-$}
\put(100,-15){$\sbottom_L$}
\put(40,32){$\stopx_L$}
\put(116,32){$\sgl$}
\CrossedCircle{75}{0}{72}{-3}{78}{3}
\put(-23,-3){+}
\put(173,-3){+}
\end{picture}
\begin{picture}(200,110)(-50,0)
\ArrowLine(0,0)(50,0)
\Line(50,0)(100,0)
\Photon(50,0)(100,0){5}{3}
\DashLine(100,0)(150,0){5}
\ArrowLine(100,50)(100,25)
\ArrowLine(100,25)(100,0)
\ArrowLine(150,0)(200,0)
\CArc(100,0)(50,0,90)
\GlueArc(100,0)(50,0,90){5}{5}
\DashCArc(100,0)(50,90,180){5}
\put(2,6){$b_R$}
\put(148,6){$b_L$}
\put(86,16){$t$}
\put(37,-18){$\tilde H_1^-$}
\put(67,-18){$\tilde W^-$}
\put(100,-15){$\sbottom_L$}
\put(30,20){$\stopx_L$}
\put(55,43){$\stopx_R$}
\put(116,32){$\sgl$}
\CrossedCircle{65}{35}{62}{32}{68}{38}
\CrossedCircle{75}{0}{72}{-3}{78}{3}
\CrossedCircle{100}{25}{97}{22}{103}{28}
\end{picture} \\
\caption{\it All possible cross insertions into one of the diagrams
contributing to the leading terms of $\Delta_{b,2}^{elw,2}$ at NNLO
involving top quarks and stops.}
\label{fg:crosses}
\end{figure}

For the finite result of the two-loop corrections to
$\Delta_{b,2}^{elw,1/2}$ we have to renormalize the sbottom and stop
masses as well as the sbottom and stop mixing angles. The counterterm
of the $\Delta_{b,2}^{elw,1}$ term can be derived as
\bea
\delta \Delta_{b,2,NNLO}^{elw,1} = \sum_{i=1,2} \frac{\partial
\Delta_{b,2}^{elw,1}}{\partial (m_{\tilde b_i}^2)} \delta m_{\tilde
b_i}^2 + \frac{\partial \Delta_{b,2}^{elw,1}}{\partial \theta_b} \delta
\theta_b
\eea
where $\Delta_{b,2}^{elw,1}$ is the NLO expression of
Eq.~(\ref{eq:effparb}).  The renormalization constants are given by
\bea
\delta\theta_b & = & -\frac{C_F}{4} \frac{\alpha_s}{\pi} \Re e\left\{
s_{2\theta_b} c_{2\theta_b} \frac{A_0(m_{\tilde b_2})-A_0(m_{\tilde
b_1})}{m_{\tilde b_2}^2 - m_{\tilde b_1}^2} \right\} \nonumber \\
\delta m_{\tilde b_i}^2 & = & \frac{C_F}{4} \frac{\alpha_s}{\pi} \Re e
\left\{ (1+c_{2\theta_b}^2) A_0(m_{\tilde b_i}) + s_{2\theta_b}^2
A_0(m_{\tilde b_j}) - 2A_0(M_{\sgl}) \right. \nonumber \\
& & \left. -4 m_{\tilde b_i}^2 B_0(m_{\tilde
b_i}^2;0,m_{\tilde b_i}) + 2(m_{\tilde b_i}^2 - M_{\sgl}^2) B_0(m_{\tilde
b_i}^2; M_{\sgl},0) \right\} \qquad (j\neq i)
\label{eq:renb}
\eea
where we neglected terms of ${\cal O}(m_b)$ consistently.

For the $\Delta_{b,2}^{elw,2}$ term we have to renormalize the stop
contributions in addition so that the full counterterm is given by
\bea
\delta \Delta_{b,2,NNLO}^{elw,2} = \sum_{i=1,2} \frac{\partial
\Delta_{b,2}^{elw,2}}{\partial (m_{\tilde b_i}^2)} \delta m_{\tilde
b_i}^2 + \frac{\partial \Delta_{b,2}^{elw,2}}{\partial \theta_b} \delta
\theta_b + \sum_{i=1,2} \frac{\partial
\Delta_{b,2}^{elw,2}}{\partial (m_{\tilde t_i}^2)} \delta m_{\tilde
t_i}^2 + \frac{\partial \Delta_{b,2}^{elw,2}}{\partial \theta_t} \delta
\theta_t
\eea
with the NLO contribution $\Delta_{b,2}^{elw,2}$ of
Eq.~(\ref{eq:effparb}), the renormalization constants $\delta m_{\tilde
b_i}^2, \delta\theta_b$ of Eq.~(\ref{eq:renb}) and
\bea
\delta\theta_t & = & \!\frac{C_F}{4} \frac{\alpha_s}{\pi} c_{2\theta_t}
\Re e\left\{
s_{2\theta_t} \frac{A_0(m_{\tilde t_2})-A_0(m_{\tilde
t_1})}{m_{\tilde t_1}^2 - m_{\tilde t_2}^2}
+ 2M_{\sgl} m_t \frac{B_0(m_{\tilde
t_1}^2;M_{\sgl},m_t) + B_0(m_{\tilde t_2}^2;M_{\sgl},m_t)}{m_{\tilde t_1}^2
- m_{\tilde t_2}^2} \right\} \nonumber \\
\delta m_{\tilde t_i}^2 & = & \frac{C_F}{4} \frac{\alpha_s}{\pi} \Re e
\left\{ (1+c_{2\theta_t}^2) A_0(m_{\tilde t_i}) + s_{2\theta_t}^2
A_0(m_{\tilde t_j}) - 2A_0(M_{\sgl}) - 2A_0(m_t) \right. \nonumber \\
& - & \left. 4 m_{\tilde t_i}^2 B_0(m_{\tilde
t_i}^2;0,m_{\tilde t_i}) - 2(M_{\sgl}^2 + m_t^2 - m_{\tilde t_i}^2 \mp
M_{\sgl} m_t s_{2\theta_t}) B_0(m_{\tilde b_i}^2; M_{\sgl},m_t) \right\}
\quad (j\neq i)
\eea
for the stop renormalization constants, where we kept all terms
proportional to the top mass $m_t$. The final results have been
explicitly checked to be ultraviolet finite after renormalization. The
whole calculation has been performed twice independently with different
methods and implementations.

Due to the mismatch between the $(n-2)$ gluonic d.o.f.~and the 2
d.o.f.~of the gluinos in dimensional regularization, anomalous
counterterms have to be added to restore supersymmetry \cite{susyrest}.
This affects the SUSY-counterparts $\hat g, \hat g'$ of the electroweak
gauge couplings at the $\tilde B q\squark$ and $\tilde W q\squark$
vertices ($q = t,b$ and $\squark = \sbottom, \stopx$),
\bea
\hat g = g \left[ 1-\frac{C_F}{8} \frac{\alpha_s}{\pi}\right], \qquad
\hat g' = g' \left[ 1-\frac{C_F}{8} \frac{\alpha_s}{\pi}\right]
\eea
as well as the SUSY-counterparts of the Higgs Yukawa
couplings\footnote{The same anomalous counterterms arise for the charged
higgsino coupling $\lambda_{H^\pm \stopx b}$, too.},
\bea
\lambda_{\tilde H q \squark} = \lambda_{Hqq} \left[ 1-\frac{3}{8} C_F
\frac{\alpha_s}{\pi}\right] , \qquad
\lambda_{H \squark \squark} = \lambda_{Hqq} \left[ 1-\frac{C_F}{4}
\frac{\alpha_s}{\pi}\right]
\eea
This results in anomalous counterterms of the contributions
$\Delta_{b,2}^{elw,1/2}$,
\bea
\delta \Delta_{b,2,anom}^{elw,1} & = & - \frac{C_F}{2} \frac{\alpha_s}{\pi}
\Delta_{b,2}^{elw,1} \nonumber \\
\delta \Delta_{b,2,anom}^{elw,2} & = & - \frac{C_F}{2} \frac{\alpha_s}{\pi}
\Delta_{b,2}^{elw,2}
\eea
where $\Delta_{b,2}^{elw,1/2}$ denote the one-loop expressions of
Eq.~(\ref{eq:effparb}).

\subsection{Strange Yukawa Couplings}
The translation of the results for the bottom-Yukawa couplings to the
Higgs boson couplings to strange quarks requires a careful investigation
of the corresponding quark-mass contributions.  Since in the calculation
of the bottom-Yukawa coupling the bottom quark is treated strictly
massless and the external momentum dependence is omitted, there is no
difference for the individual two-loop diagrams, if the bottom
parameters are replaced by their corresponding strange parameters. Care
must be taken for the proper summation over all quark/squark flavours
for the diagrams with gluino-self-energy insertions since the
strange-squark mass coincides with the left- and right-handed squark
masses of the first two generations and the sbottom and stop masses of
the third generation are independent. Another difference to the
bottom-quark case is the absence of sizeable charm-Yukawa-induced
SUSY--electroweak contributions to the strange-Yukawa coupling, since we
are neglecting the charm Yukawa coupling $\lambda_c$. The final result
can be cast into the form
\bea
\Delta_s = \frac{\Delta_{s,2}^{QCD} \left[ 1+\delta_s \right] +
\Delta_{s,2}^{elw,1} \left[ 1+\delta_1 \right] + \Delta_{s,2}^{elw,2}
\left[ 1+\delta_2 \right]} {1+\Delta_{s,1} \left[ 1+\delta_s
\right]}~\tgb
\label{eq:deltas}
\eea
where $\delta_s$ ($\delta_{1/2}$) denotes the NNLO SUSY--QCD corrections
to the QCD (electroweak) part of the strange Yukawa couplings and
$\Delta^{QCD}_{s,1/2}, \Delta^{elw, 1/2}_{s,2}$ are defined in
Eq.~(\ref{eq:effpars}). The expression above for $\Delta_s$ is then
inserted into the effective Lagrangian of Eq.~(\ref{eq:leff}) and into the
resummed couplings of Eq.~(\ref{eq:rescoup}), respectively, resumming in
this way all terms of ${\cal O}\left[(\Delta_s)^n\right]$.

\section{Results}
The final results of this paper have been implemented in the program
{\tt Hdecay} \cite{hdecay}. This code computes the MSSM Higgs couplings and
masses based on the RG-improved expressions of Ref.~\cite{rgi}.
Moreover, the partial decay widths and branching ratios of the MSSM
Higgs bosons are calculated including higher-order corrections
\cite{review}. For large $\tgb$ values the dominant neutral Higgs boson
decays are into $b\bar b$ and $\tau^+\tau^-$. In Ref.~\cite{GHS} their
branching ratios have been analyzed including the correction $\Delta_b$
of Eq.~(\ref{eq:effparb}) at the one-loop level.

\subsection{Higgs Decays into Bottom and Strange Quarks}
The QCD and SUSY--QCD corrected partial decay widths of the neutral
Higgs bosons $\Phi=h,H,A$ into bottom quarks can be expressed as
\cite{GHS}
\begin{equation}
\Gamma [\Phi \, \to \, b{\overline{b}}] =
\frac{3G_F M_\Phi }{4\sqrt{2}\pi} \overline{m}_b^2(M_\Phi)
\left[ 1 + \delta_{\rm QCD} + \delta_t^\Phi \right] \tilde g_b^\Phi \left[
\tilde g_b^\Phi + \delta_{SQCD}^{rem} \right]
\label{eq:gambb}
\end{equation}
with $\overline{m}_b(M_\Phi)$ denoting the $\overline{\rm MS}$ bottom
mass at the scale of the Higgs mass $M_\Phi$ and quark mass effects
beyond ${\cal O}(m_b^2)$ are neglected. The QCD corrections $\delta_{\rm
QCD}$ and the top quark induced contributions $\delta_t^\Phi$ are known
\cite{drees} and can be found in Ref.~\cite{review} in compact form. The
QCD corrections $\delta_{\rm QCD}$ are taken into account up to N$^4$LO
and the corrections $\delta_t^\Phi$ at the NNLO level in {\tt Hdecay}.

The leading contributions of the SUSY--QCD corrections \cite{solaeberl}
have been absorbed in the resummed bottom-Yukawa couplings $\tilde
g_b^\phi$ as given in Eq.~(\ref{eq:rescoup}). The remainder
$\delta_{SQCD}^{rem}$ is small, i.e.~at the sub-per-cent level, in
phenomenologically relevant scenarios for large $\tgb$ values
\cite{GHS}. This observation at NLO implies the expectation that at
higher orders the remainders after factorizing the corrections involved
in the effective Lagrangian of Eq.~(\ref{eq:leff2}) are even smaller and
thus negligible in general. This underlines that the results of our work
constitute the major part of the corrections beyond NLO with tiny
remainders at higher loop-levels. It should be noted that our two-loop
corrections to the $A_b,A_s$-induced terms are formally of three-loop
order of the related physical observables involving the corresponding
effective Yukawa couplings, but they modify the relations between the
Yukawa couplings and quark masses at NNLO. In our analysis we include
the full (two-loop corrected) $\Delta_b$ ($\Delta_s$) contributions
including the QCD and electroweak parts in the couplings $\tilde
g_b^\phi$ ($\tilde g_s^\phi$).

The strange Yukawa coupling plays a phenomenological role for charged
Higgs decays into charm and strange quarks $H^+\to c\bar s$. Neglecting
regular quark mass effects\footnote{{\tt Hdecay} includes the full quark-mass
dependence up to NLO.} this partial decay width can be expressed as
\cite{h2cs}\footnote{$V_{cs}$ denotes the corresponding CKM-matrix
element.}
\begin{equation}
\Gamma [\,H^{+} \to \, c{\overline{s}}\,] =
\frac{3 G_F M_{H^\pm}}{4\sqrt{2}\pi} \, \left| V_{cs} \right|^2 \,
\left[ \overline{m}_c^2(M_{H^\pm}) (g_c^{A})^2 +
\overline{m}_s^2(M_{H^\pm})
 (\tilde g_s^{A})^2 \right] (1 + \delta_{\rm QCD})
\label{eq:hcud}
\end{equation}
with the same QCD-correction-factor $\delta_{QCD}$ as in
Eq.~(\ref{eq:gambb}). The small remainder of the genuine SUSY--QCD
corrections after absorbing the dominant part in the effective
strange-Yukawa coupling $\tilde g_s^{A}$ is neglected.

\subsection{Numerical Results}
We perform our numerical analysis of the MSSM Higgs boson decays into
bottom and strange quarks for the MSSM benchmark scenario $M_h^{125}$
\cite{bench} as a representative case\footnote{The values for $A_b,
A_\tau, A_t$ have been obtained from $X_t = A_t - \mu/\tgb = 2.8$ TeV.
The trilinear coupling $A_s$ has been chosen as $A_s=A_b$.  The soft
SUSY-breaking squark mass parameter $M_{\tilde Q}$ is defined in the
on-shell scheme. We have determined the corresponding $\overline{\rm
MS}$ parameters by appropriate iterations. The $\overline{\rm MS}$
squark-mass parameters of the second generation have been identified
with the corresponding ones of the third generation.}:
\begin{eqnarray}
\mbox{$M_h^{125}$:} &&\tb = 40,\quad M_{\tilde Q} = 1.5~{\rm TeV},\quad
M_{\tilde \ell_3} = 2~{\rm TeV},\quad M_{\gluino} = 2.5~{\rm TeV},
\nonumber \\
&& M_1 = M_2 = 1~{\rm TeV},\quad A_b = A_\tau = A_t = 2.825~{\rm TeV},\quad
\mu = 1~{\rm TeV}
\end{eqnarray}
For the Higgs masses and couplings we use the RG-improved two-loop
expressions of Ref.~\cite{rgi}, so that the leading corrections at the
one- and two-loop level to the Higgs masses and the effective mixing
angle $\alpha$ are taken into account.  The bottom $\overline{\rm
MS}$-mass has been chosen as $\overline{m_b}(\overline{m_b})=4.18$~GeV,
which corresponds to a pole mass $m_b=4.84$~GeV according to the
implementation in {\tt Hdecay} that determines the bottom-quark pole mass from
the $\overline{\rm MS}$-mass at the scale of the pole mass. The strange
$\overline{\rm MS}$-mass has been initialized as $\overline{m_s}(2~{\rm
GeV})=95$~MeV and the strong coupling constant as $\alpha_s(M_Z)=0.118$.
The top pole mass has been taken to be $m_t=172.5$ GeV. In the mass
matrices of the stop and sbottom states effective top and bottom masses
are implemented as discussed in Ref.~\cite{hsqsq}. With this set-up we
obtain the following squark masses e.g.~for $\tgb=40$,
\bea
m_{\stopx_1} \!\!\! & = & \!\!\! 1374.8~{\rm GeV}, \quad m_{\stopx_2} =
1632.6~{\rm GeV}, \quad m_{\sbottom_1} = 1479.3~{\rm GeV}, \quad
m_{\sbottom_2} = 1525.3~{\rm GeV}, \nonumber \\
m_{\tilde c_1} \!\!\! & = & \!\!\! 1502.6~{\rm GeV}, \quad m_{\tilde
c_2} = 1503.1~{\rm GeV}, \quad m_{\tilde s_1} = 1500.2~{\rm GeV}, \quad
m_{\tilde s_2} = 1504.8~{\rm GeV}
\eea

\begin{figure}
\begin{center}
\vspace*{-3.0cm}
\epsfig{file=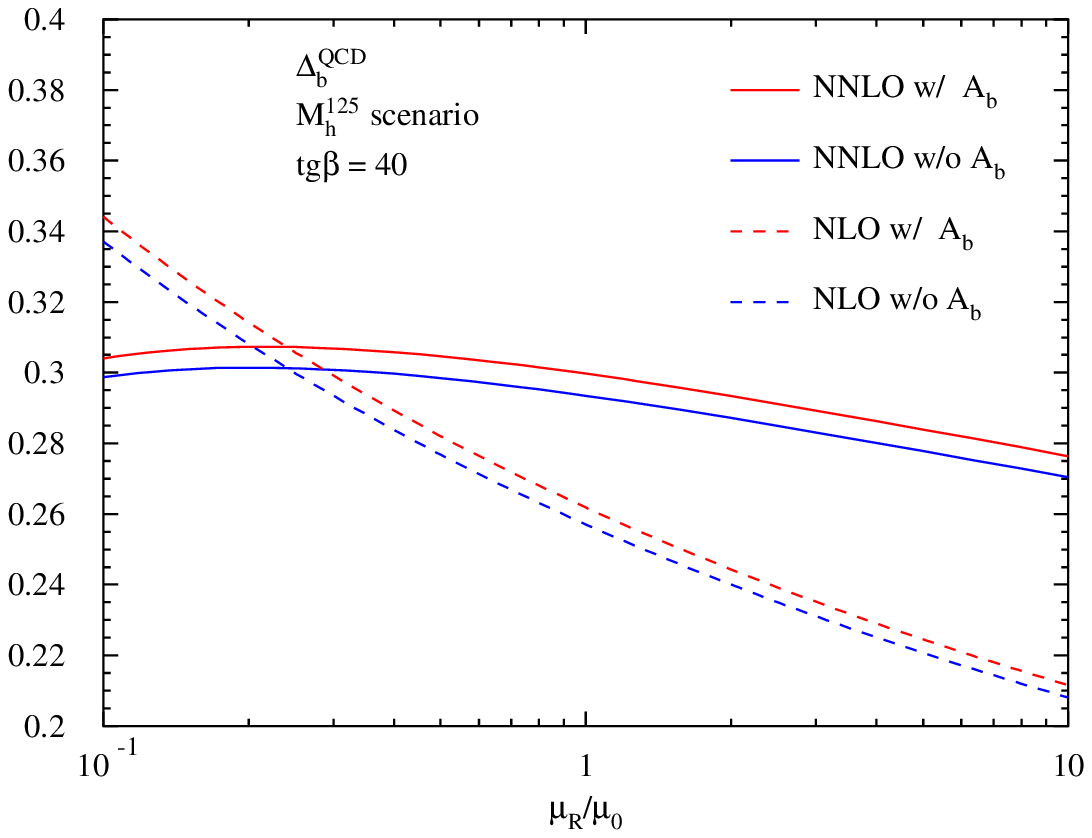,%
        bbllx=30pt,bblly=350pt,bburx=520pt,bbury=650pt,%
        scale=0.6}
\vspace*{2.5cm}

\epsfig{file=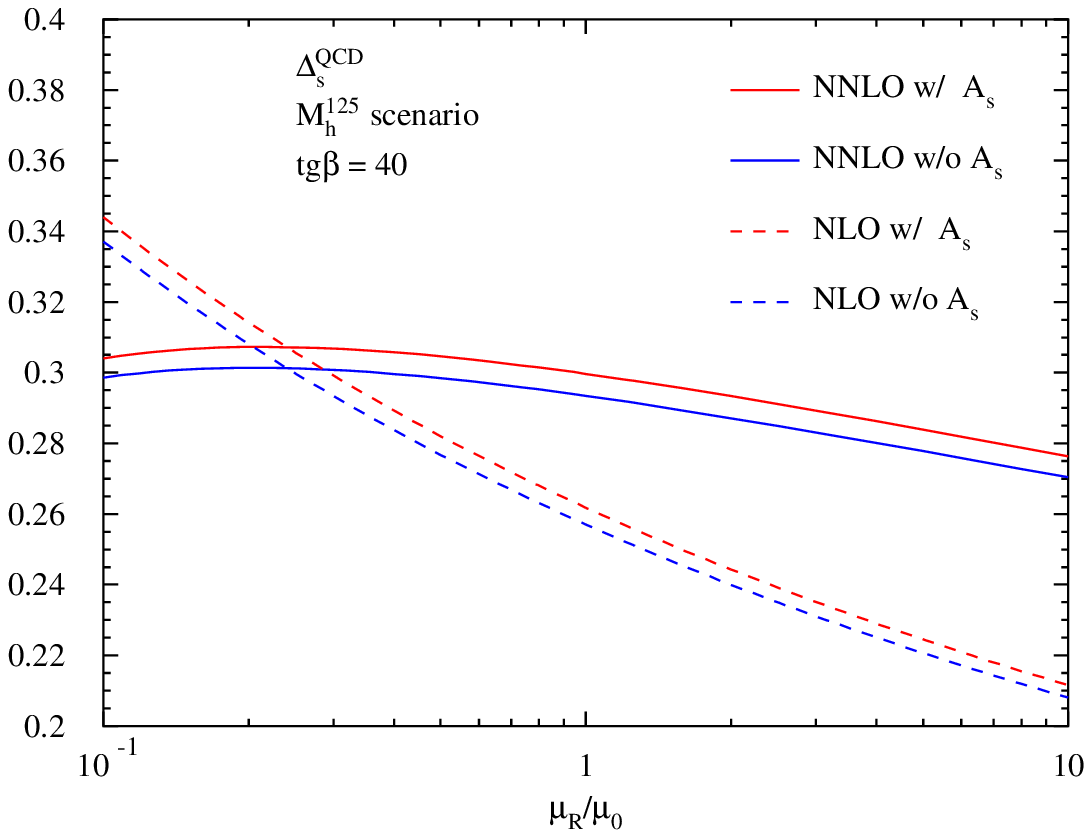,%
        bbllx=30pt,bblly=350pt,bburx=520pt,bbury=650pt,%
        scale=0.6}
\end{center}
\vspace*{2.0cm}
\caption{\it Scale dependence of the SUSY--QCD corrections
$\Delta_b^{QCD}$ (upper plot) and $\Delta_s^{QCD}$ (lower plot) at
one-loop and two-loop order with and without the $A_b,A_s$ contributions
in the $M_h^{125}$ scenario.}
\label{fg:scale_mh125}
\end{figure}

We will present the impact of the new results on the bottom- and
strange-Yukawa couplings as well as related observables.
Fig.~\ref{fg:scale_mh125} displays the scale dependence of the SUSY--QCD
parts of $\Delta_b$ and $\Delta_s$,
\bea
\Delta_q^{QCD} = \frac{\Delta_{q,2}^{QCD} (1+\delta_q)}{1+\Delta_{q,1}
(1+\delta_q)}~\tgb \qquad (q=b,s)
\eea
with and without the $A_b,A_s$ contributions $\Delta_{b,1},
\Delta_{s,1}$ for the $M_h^{125}$ scenario. For $\tgb=40$ the $\Delta_b$
and $\Delta_s$ corrections range at the level of 30\%.  The scale
dependence is reduced significantly from one- to two-loop order to a few
per cent at NNLO, while the additional contributions of the $A_b, A_s$
terms are small, i.e.~at the per-cent level, as can be inferred from the
differences between the red and blue curves. The effects of the
contributions induced by $A_b, A_s$ are significantly larger than the
NLO remainder so that this can naturally be expected for the remainders
at NNLO and beyond, too.  However, in general the sizes and directions
of the total $\Delta_b$ and $\Delta_s$ corrections depend on the MSSM
scenario, in particular on the sign and size of $\mu$ and the value of
$\tgb$. The central scales have been chosen equal to the average of the
corresponding SUSY masses, i.e.~$\mu_0=(m_{\squark_1} + m_{\squark_2} +
m_{\sgl})/3$ for $\Delta_{b,1/2}^{QCD}$, $\Delta_{s,1/2}^{QCD}$. For the
electroweak parts we have adopted $\mu_0=(m_{\st_1} + m_{\st_2} +
\mu)/3$ for $\Delta_{b,2}^{elw,t}$ and the average of the involved SUSY
particles for the strong-coupling scale involved in the individual
contributions to $\Delta_{q,2}^{elw,1/2}$ $(q=b,s)$ at NNLO. These
choices are suitable as the appropriate natural central scales.

\begin{figure}
\begin{center}
\vspace*{-3.0cm}
\epsfig{file=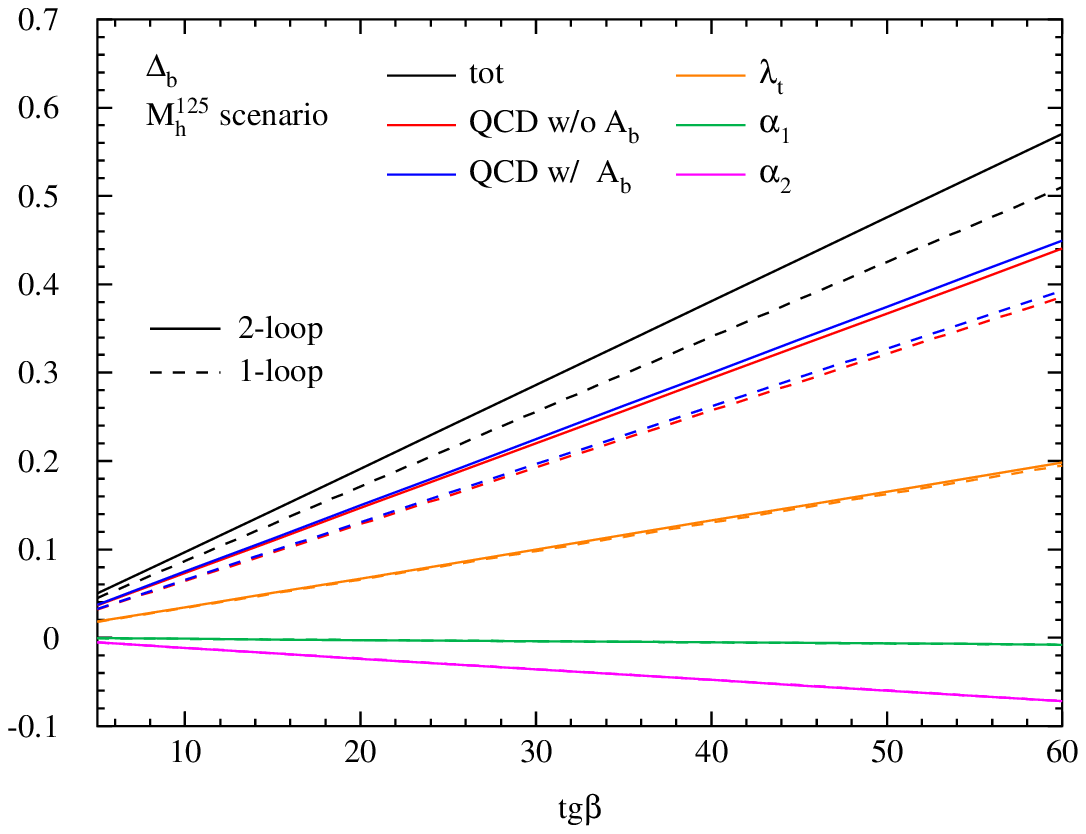,%
        bbllx=30pt,bblly=350pt,bburx=520pt,bbury=650pt,%
        scale=0.6}
\vspace*{2.5cm}

\epsfig{file=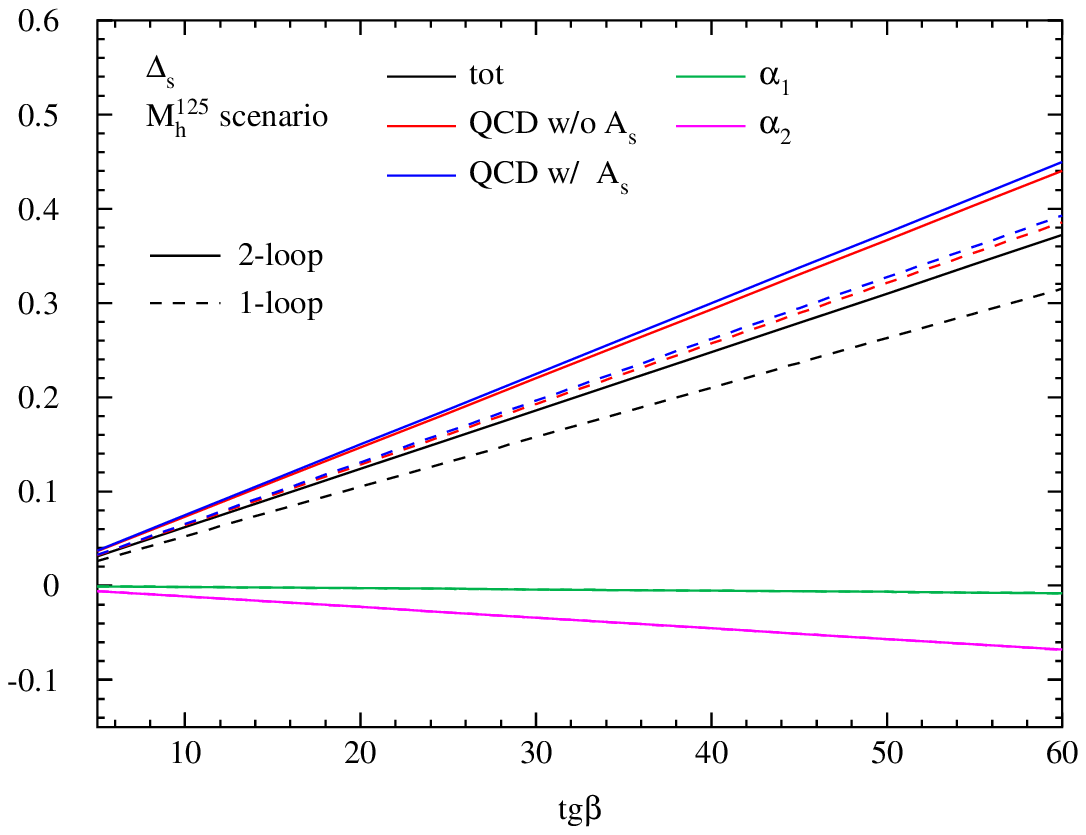,%
        bbllx=30pt,bblly=350pt,bburx=520pt,bbury=650pt,%
        scale=0.6}
\end{center}
\vspace*{2.0cm}
\caption{\it Dependence of the full SUSY--QCD + SUSY--electroweak
corrections $\Delta_b$ (upper plot) and $\Delta_s$ (lower plot) on
$\tgb$ at one-loop and two-loop order with and without the $A_b,A_s$
contributions in the $M_h^{125}$ scenario, and the individual
contributions to $\Delta_b$ and $\Delta_s$.}
\label{fg:tgb}
\end{figure}

In Fig.~\ref{fg:tgb} we present the individual and total contributions
to $\Delta_b$ and $\Delta_s$ as a function of $\tgb$ for the $M_h^{125}$
scenario\footnote{The values of $A_t=A_b=A_\tau$ have been derived from
$X_t=2.8$ TeV for each value of $\tgb$ accordingly.}. While the SUSY-QCD
parts $\Delta_{b/s,2}^{QCD}$ dominate the contributions, the electroweak
contributions $\Delta_{b/s,2}^{elw,t/1/2}$ amount to ${\cal O}(10\%)$
reaching up to about 20\% for large $\tgb$-values. Especially the terms
$\Delta_{b/s,2}^{elw,t/2}$ cannot be neglected if a prediction with an
accuracy at the per-cent level should be achieved. The two-loop
corrections to these terms range at the per-cent level in this benchmark
scenario. They can be larger in other scenarios.

\begin{figure}[hbtp]
\begin{center}
\vspace*{-0.3cm}

\hspace*{-1.3cm}
\epsfig{file=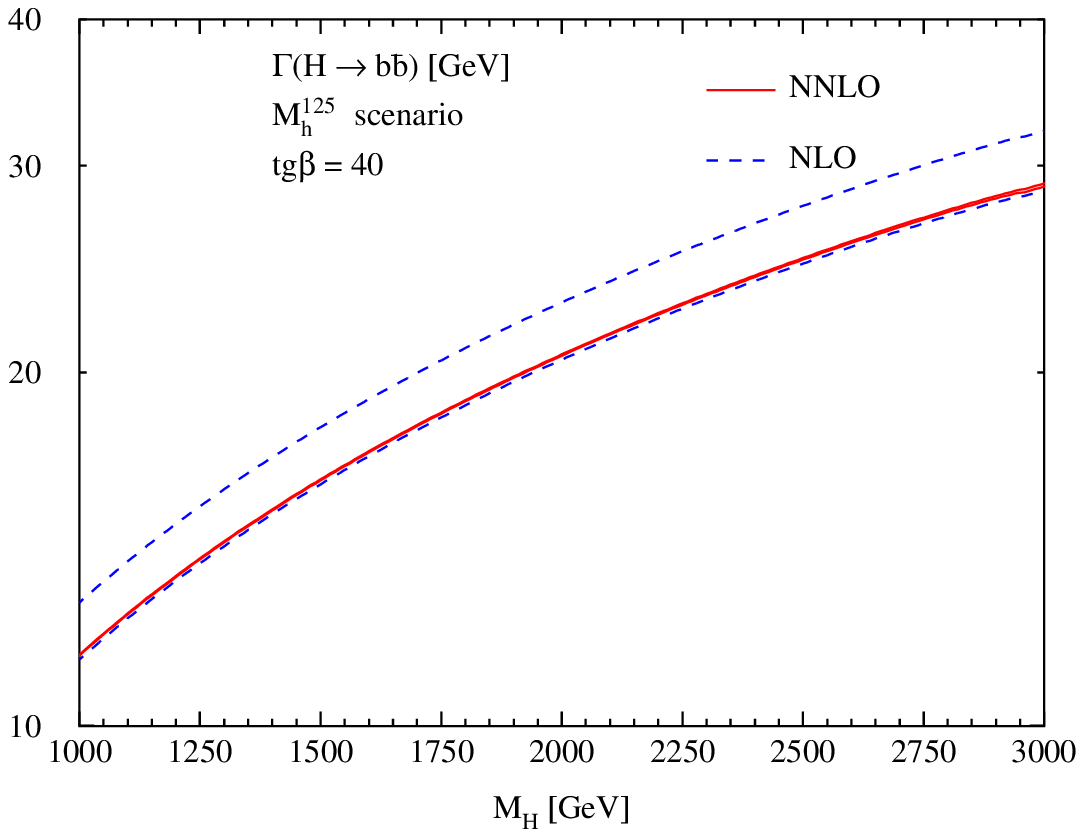,%
        bbllx=30pt,bblly=350pt,bburx=520pt,bbury=650pt,%
        scale=0.45}
\vspace*{1.68cm}

\hspace*{-1.3cm}
\epsfig{file=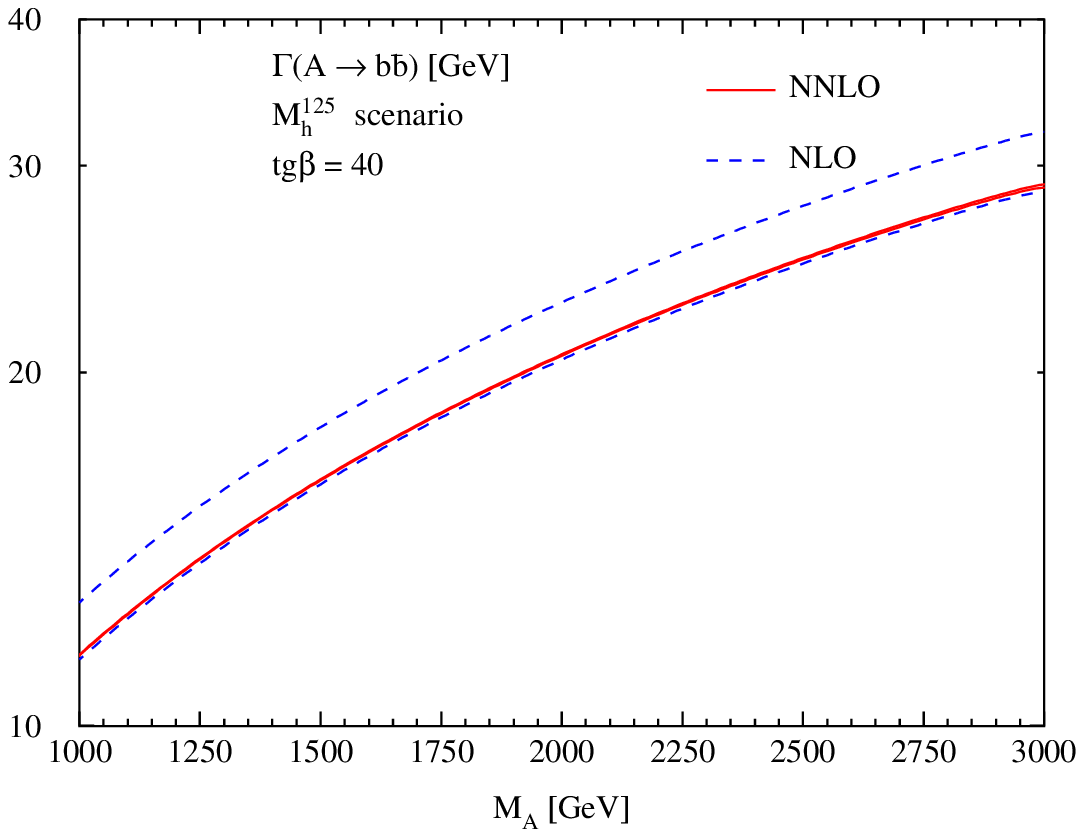,%
        bbllx=30pt,bblly=350pt,bburx=520pt,bbury=650pt,%
        scale=0.45}
\vspace*{1.68cm}

\hspace*{-1.3cm}
\epsfig{file=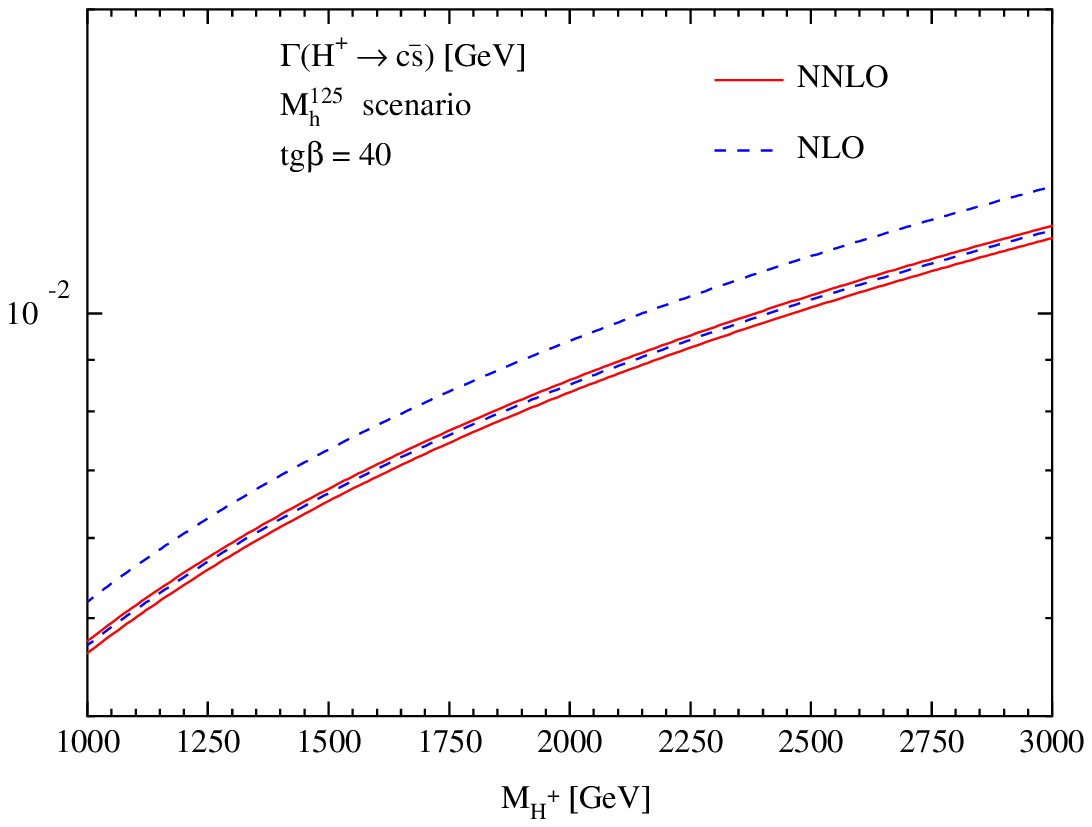,%
        bbllx=30pt,bblly=350pt,bburx=520pt,bbury=650pt,%
        scale=0.45}
\end{center}
\vspace*{1.19cm}
\caption{\it Partial decay widths of the heavy scalar $H$ and the
pseudoscalar $A$ Higgs bosons to $b\bar b$ (upper two plots) and charged
Higgs decays to $c\bar s$ (lower plot) in the $M_h^{125}$ scenario.
The dashed blue bands display the scale dependence at the one-loop level
and the full red bands at the two-loop level by varying the renormalization
scales of $\Delta_b$ and $\Delta_s$ between $1/3$ and $3$ times the
central scale fixed by the average of the involved SUSY-particle
masses.}
\label{fg:Gamma}
\end{figure}
As a particular application we analyze the partial decay widths of the
heavy neutral MSSM Higgs bosons into $b\bar b$ pairs and of the charged
Higgs boson into $c\bar s$ in Fig.~\ref{fg:Gamma} for the $M_h^{125}$
scenario, respectively. The two-loop parts of $\Delta_b$ ($\Delta_s$)
reduce the partial decay widths to $b\bar b$ ($c\bar s$) pairs for the
central scale choices by ${\cal O}(10\%)$. The NLO bands (dashed blue
curves) and the NNLO bands (full red curves) are generated by varying
the renormalization scales of $\Delta_b$ and $\Delta_s$ between $1/3$
and $3$ times the corresponding central scales $\mu_0$\footnote{We have
extended the usual range to a factor of three, because for a factor of
two, the bands do not overlap so that the more restricted range is not
appropriate.}.  A significant reduction of the dashed one-loop bands of
${\cal O}(10\%)$ to the full two-loop bands at the per-cent level can be
inferred from these results which is expected based on the previously
known two-loop corrections \cite{deltabnnlo}. All NNLO results are
positioned at the lower ends of the NLO error bands.

\section{Conclusions}
We have calculated the NNLO corrections to the effective bottom- and
strange-quark-Yukawa couplings within the MSSM, extending previous
analyses to non-leading terms that are mediated by the soft
SUSY-breaking trilinear couplings $A_b,A_s$ and the weak couplings
$\alpha_1, \alpha_2$ for large values of $\tgb$.  The dominant
contributions of the SUSY--QCD corrections arise from virtual two-loop
squark and gluino exchange, that factorize and can be absorbed in
effective Yukawa couplings.  We have derived the two-loop SUSY--QCD
corrections to the effective bottom- and strange-Yukawa couplings beyond
the leading contributions obtained previously.

In summary, the significant scale dependence of $\Order{10\%}$ of the
NLO predictions for processes involving the bottom- and
strange-quark-Yukawa couplings of MSSM Higgs bosons necessitate the
inclusion of NNLO corrections. For the NNLO-corrected Yukawa couplings,
we observe a reduction of the scale dependence to the per-cent level.
These results were previously known for the leading terms of the
bottom-Yukawa couplings. In this work they have been established for the
non-leading $A_b$ and electroweak terms and the strange-Yukawa
couplings.  The improved NNLO results for the bottom- and strange-Yukawa
couplings provide a quantitative basis for experimental analyses at the
LHC and future $e^+e^-$ colliders as the ILC. \\

\noindent
{\bf Acknowledgments} \\
The authors are indebted to M.~M\"uhlleitner for carefully reading the
manuscript and very useful comments. This work is supported in part by
the Swiss National Science Foundation.



\end{document}